\documentclass{article}

\usepackage[left = 20 mm, right = 20 mm, top = 30 mm, bottom = 20 mm]{geometry}
\usepackage{amsmath, amssymb, verbatim, graphicx, subfig, caption, slashed, empheq, cite, jheppub}
\usepackage[toc,page]{appendix}

\newcommand{\minus}{\scalebox{0.75}[1.0]{$-$}}

\allowdisplaybreaks

\begin{document}

\title{Generation of quasiparticles by flavor mixing and CP violation:
heavy Majorana neutrinos, part I}

\author{Chang-Hun Lee}
\affiliation{National Center for Theoretical Sciences, \\
101, Section 2, Kuang-Fu Road, Hsinchu, \\
Republic of China (Taiwan)}
\emailAdd{chlee@cts.nthu.edu.tw}

\begin{abstract}
{The phenomenology of flavor mixing of heavy Majorana neutrinos is studied. The physical degrees of freedom, which propagate like free particles until they decay in the presence of flavor mixing, are identified by diagonalizing the resummed propagator. It is shown that they should be interpreted as quasiparticles which lose Majorana nature.}
\end{abstract}

\maketitle

\section{Introduction}		\label{sec:Intro}
The Standard Model of particle physics does not explain the non-zero masses of neutrinos, and one of the simplest ways to solve the problem is introducing heavy right-handed (RH) Majorana neutrinos into the model, which can generate small non-zero masses of light neutrinos by the seesaw mechanism \cite{LNC, NLRS, LRSVP}. Moreover, those heavy Majorana neutrinos can simultaneously solve another important problem of particle physics, \textit{i.e.}, the origin of matter. The CP violation effect in the decay of heavy Majorana neutrinos can generate an asymmetry between matter and antimatter, which can explain the current matter density observed in the universe. This mechanism is called leptogenesis \cite{Lepto}.

In order to generate such a CP violation effect, the presence of multiple flavors of Majorana neutrinos is essential. When the masses of heavy Majorana neutrinos are almost degenerate, the CP asymmetry generated by mixing of those multiple flavors can be hugely enhanced such that even masses of heavy Majorana neutrinos in the TeV scale are large enough to allow successful leptogenesis. This mechanism is called resonant leptogenesis \cite{CPVMaj, ResLepto}, and this scenario is particularly interesting since such relatively light RH Majorana neutrinos can be discovered in the particle colliders currently available.

\begin{figure}[h]
	\centering
	\subfloat[]{
		\includegraphics[width = 27 mm]{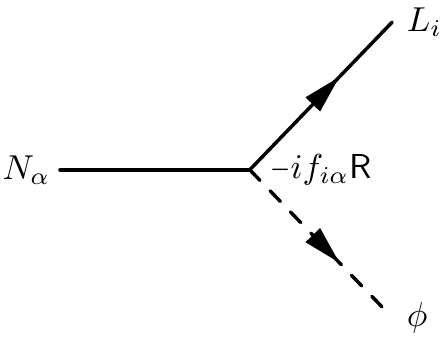}
	} \qquad
	\subfloat[]{
		\includegraphics[width = 40 mm]{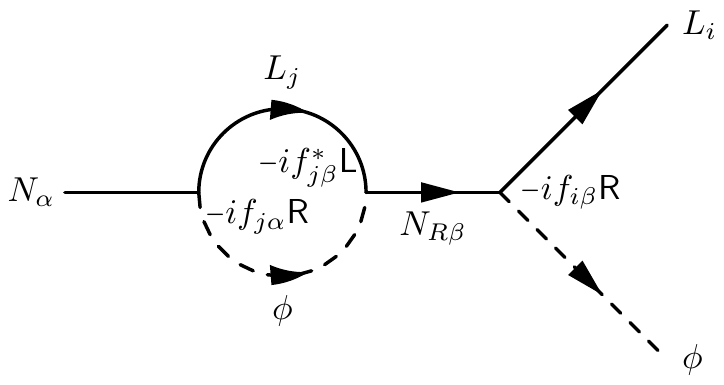}
	} \qquad
	\subfloat[]{
		\includegraphics[width = 40 mm]{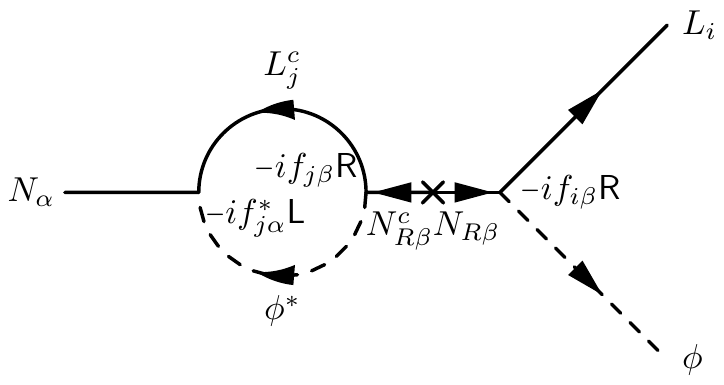}
	} \\
	\subfloat[]{
		\includegraphics[width = 27 mm]{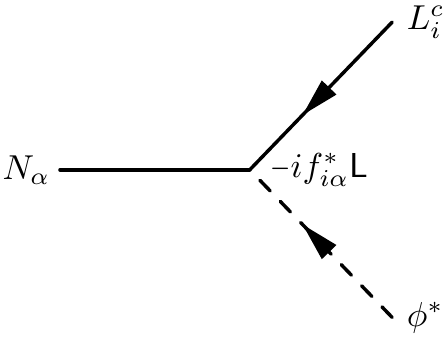}
	} \qquad
	\subfloat[]{
		\includegraphics[width = 40 mm]{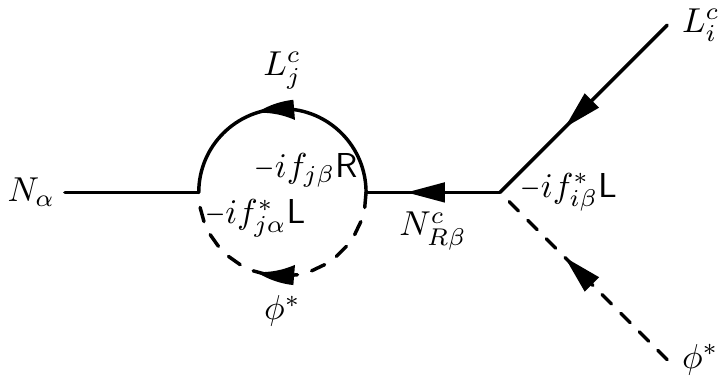}
	} \qquad
	\subfloat[]{
		\includegraphics[width = 40 mm]{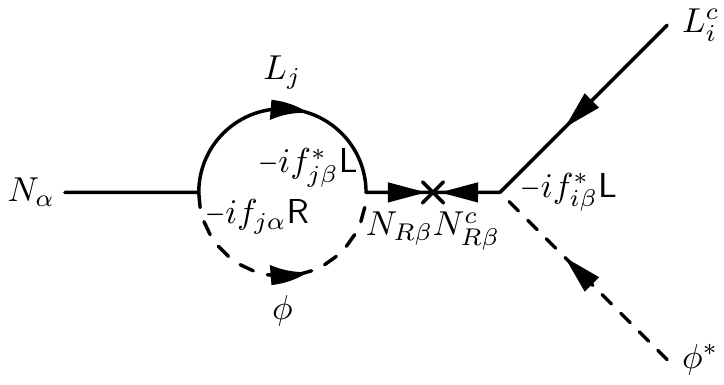}
	}
	\caption{Tree-level diagrams and their one-loop corrections to the decaying heavy Majorana neutrinos used in references \cite{CPVLepto, CPMajDec}. This approach is inappropriate because of the divergences in the diagrams with loops.}
	\label{fig:N1Loop}
\end{figure}

Hence, the theoretical analysis of flavor mixing and CP violation has been an interesting research topic. The first expression of the CP asymmetry in the decays of heavy Majorana neutrinos was derived in reference \cite{CPVLepto}, where it was obtained from non-amputated diagrams with one-loop corrections to the decaying heavy Majorana fields. The associated Feynman diagrams are given in figure \ref{fig:N1Loop}, where $N_\alpha$, $L_i$, and $\phi$ denote the heavy Majorana field, left-handed (LH) lepton SU(2)-doublet, and Higgs SU(2)-doublet, respectively.
The resulting expression is, however, divergent when the masses of heavy neutrinos are degenerate. In order to find an expression of the CP asymmetry with a regulator such that it is applicable to resonant leptogenesis, several different methods were used. In reference \cite{CPVMaj}, it was obtained from the same non-amputated diagrams with the tree-level internal propagators of Majorana neutrinos replaced by sorts of resummed propagators. Alternatively in reference \cite{ResLepto}, the CP asymmetry is obtained by expanding the non-diagonal resummed propagator around its poles, and an identical expression to that in reference \cite{CPVMaj} was obtained. However, in references \cite{CPMajDec, CPResLepto}, the CP asymmetry was calculated by diagonalizing the resummed propagator matrix of heavy Majorana neutrinos, and a different regulator was derived. In a different approach discussed in references \cite{LeptoFPRes, KBResLepto}, the CP asymmetry is calculated in the framework of the non-equilibrium quantum field theory, and another form of the regulator was obtained.

In fact, there exist some problems in the derivations given in the literature. The way how the CP asymmetry was calculated in reference \cite{CPVLepto} is actually flawed since it is inappropriate to consider the loop correction to heavy neutrino fields only up to the single one-particle irreducible (1PI) contributions as in figure \ref{fig:N1Loop}. In that approach, the internal propagator $i / (\slashed{p} - m_{N_\beta})$ diverges for $\beta = \alpha$ since $N_\alpha$ is on-shell as an external field, \textit{i.e.}, $\slashed{p} = m_{N_\alpha}$, and such a divergence can be avoided by intentionally disregarding the contribution of $\beta = \alpha$. The only way to legitimately handle such a divergence (or non-perturbative effect) is to consider the loop corrections by resummation. Instead of considering the decays of heavy neutrinos, the loop corrections to the propagator of heavy neutrinos should be taken into account for resummation. We can resolve the problem of divergence, first resumming loop corrections outside the region of divergence and then analytically continuing it to the region of divergence in the complex plane of $p^2$. In contrast, replacing the internal propagator in the loop diagrams of figure \ref{fig:N1Loop} with a resummed propagator does not work, since the resulting $S$-matrix element vanishes for on-shell $N_\alpha$, similarly to what was shown for particle-antiparticle mixing in reference \cite{QFTMixing}. In the derivation discussed in reference \cite{CPVMaj}, the tree-level propagator $i / (\slashed{p} - m_{N_\beta})$ was replaced by $i / [\slashed{p} - m_{N_\beta} + \Sigma_{\beta \beta} (\slashed{p})]$, which implies that the resummation had not been fully taken. Hence, a non-vanishing $S$-matrix element could be erroneously obtained.

In an alternative approach in reference \cite{CPMajDec}, the pole expansion of the non-diagonal resummed propagator was performed under the assumption that the field associated with the physical pole is close to $N_\alpha$, which was thought to be achieved by on-shell renormalization as in many other works which studied renormalization of the theory with multiple flavors of Majorana fields \cite{MixRenMaj, ProbMixRen, CPVMaj, CPResLepto, ResLepto, FlavCovResLepto}. However, this is possible only when the mass difference between flavors is large, \textit{i.e.}, $\Delta m_N \gg \Gamma_N$. When the mass difference is small, \textit{i.e.}, $\Delta m_N \lesssim \Gamma_N$, the mixing matrix from $N_\alpha$ to the physical degrees of freedom requires large non-unitary mixing among flavors. Such a mixing matrix cannot be absorbed into the field-strength renormalization factor $Z_N$, since it should satisfy $Z_N = U + \mathcal{O} (f^2 / 4 \pi)$ where $U$ is a unitary matrix and $f$ is the Yukawa coupling. Otherwise, the perturbativity of the theory represented by $\mathcal{O} (f^2 / 4 \pi)$ cannot be maintained. In other words, we cannot find counterterms of $\mathcal{O} (f^2 / 4 \pi)$, if such a large non-unitary mixing matrix is absorbed into $Z_N$ so that $Z_N \neq U + \mathcal{O} (f^2 / 4 \pi)$. Hence, in the case of a small mass difference, the on-shell renormalization scheme or complex mass scheme cannot be applied, and the physical degree of freedom should be much different from $N_\alpha$. We will see that the generic non-perturbative effect in an on-shell unstable particle causes such large non-unitary mixing of flavors.

In references \cite{CPMajDec, CPResLepto}, the CP asymmetry was derived under the assumption of $\Delta m / m \gg \mathcal{O} (f^2 / 4 \pi)$. Hence, it is evident that their result cannot be applied to resonant leptogenesis in which the CP asymmetry is thought to be maximal when $\Delta m \sim \Gamma_N$. In fact, even for a large mass difference, their result cannot be said to be an improvement over the expression derived in reference \cite{CPVLepto}, since the consistency in perturbative expansion was lost in the derivation: the contributions of $\mathcal{O} (\Sigma^2)$ was neglected in calculating the effective Yukawa coupling, while the regulator which was kept is also an effect of $\mathcal{O} (\Sigma^2)$ to the whole resulting expression. This problem is actually common in the other works in the literature which aimed at obtaining the regulator, and thus it is no surprise that they obtained different regulators because neglecting the contributions of $\mathcal{O} (\Sigma^2)$ was done in different ways. We will see that it is not such a small perturbative correction which regulates the effective Yukawa couplings.

In addition, the discussions in references \cite{LeptoFPRes, KBResLepto} are based on thermal physics, as mentioned above. Even though the basic motivation of research is closely related to resonant leptogenesis which requires the thermal environment, the CP asymmetry is by itself a quantity that must be well-defined for any mass difference at zero temperature. We should be able to calculate the CP asymmetry without introducing thermal physics.

In this paper, we try to solve the problem by diagonalizing the resummed propagator and expanding it around the poles. Since the resummed propagator of heavy Majorana fields has a complicated chiral structure, its diagonalization turns out to be a difficult task. Using the associated mixing matrices, we will identify the field corresponding to each component of the diagonalized propagator, and show that it cannot be written as a single linear combination of $N_\alpha$. In consequence, it turns out that the degree of freedom that propagates like a free particle until it decays should be interpreted as a \textit{quasiparticle}, \textit{i.e.}, an emergent particle dynamically generated by interactions. Since the physics of quasiparticles requires an extensive study by itself, we will not try to provide the complete solution to the problem of finding the CP asymmetry here. The derivations of the decay widths and CP asymmetry will be discussed in a follow-up paper \cite{MajMixingII}. Even though the overall discussion will be similar to that in reference \cite{QFTMixing}, each step in the case of Majorana particles turns out to be more challenging in general. Since the on-shell renormalization scheme or complex mass scheme cannot be blindly applied to any mass difference as mentioned above, we will also carefully renormalize the Lagrangian and self-energy step-by-step, considering the constraints on the counterterms. Even though all the calculations will be done up to the one-loop order in the self-energy in this paper, it should be emphasized that such a precision does not mean that it is legitimate to calculate the loop corrections as in figure \ref{fig:N1Loop}. The generic non-perturbative effect in an on-shell unstable particle makes the collective loop effects go beyond the typical perturbative correction of the theory, and resummation is the way to handle it.

This paper is organized as follows: in section \ref{sec:BasicMaj}, some basic properties of Majorana fields are briefly reviewed; in section \ref{sec:Single}, mass and field-strength renormalization for a single flavor of a RH Majorana neutrino is examined; in section \ref{sec:Multiple}, the discussion of renormalization is generalized to the case of multiple flavors, and the renormalized resummed propagator is derived. The constraint on the renormalized self-energy is also discussed; in section \ref{sec:Diag}, the resummed propagator is diagonalized to find the pole masses and total decay widths of RH Majorana neutrinos. The expressions of the mixing matrices, self-energy, counterterms, and residue of the propagator are also derived; in section \ref{sec:Quasi}, it is shown that the degree of freedom which has the time evolution of the damped plane wave should be interpreted as a quasiparticle; in section \ref{sec:Example}, multiple examples for various mass differences are presented; in appendix \ref{sec:TrickMaj}, the trick to simplify the calculation of diagrams with Majorana propagators is explained; in appendix \ref{sec:1Loop}, various one-loop diagrams are explicitly calculated.

\section{Basic properties of Majorana fields}		\label{sec:BasicMaj}
In this section, we briefly review the properties of RH Majorana neutrinos. The renormalized Lagrangian involving RH Majorana neutrinos is written as
\begin{align}
	\mathcal{L} &= \frac{1}{2} \sum_\alpha \overline{N_\alpha} i \slashed{\partial} N_\alpha - \frac{1}{2} \sum_\alpha m_{N_\alpha} \overline{N_\alpha} N_\alpha
			- \sum_{i, \alpha} f_{i \alpha} \overline{L_i} \widetilde{\phi} \mathsf{R} N_\alpha - \sum_{i, \alpha} f^*_{i \alpha} \overline{N_\alpha} \widetilde{\phi}^\dag \mathsf{L} L_i
\end{align}
where $N_\alpha$, $L_i$, and $\phi$ are the RH Majorana neutrino, LH lepton SU(2)-doublet, and scalar SU(2)-doublet, respectively. Throughout the paper, Greek indices always denote RH neutrino flavors, while Latin indices usually denote LH lepton flavors. We have defined $\widetilde{\phi} \coloneqq i \sigma^2 \phi$ as usual, and $\mathsf{R}$, $\mathsf{L}$ are the chiral projection operators defined by $\mathsf{R} \coloneqq (1 + \gamma^5) / 2$ and $\mathsf{L} \coloneqq (1 - \gamma^5) / 2$. The parameters $m_{N_\alpha}$ and $f_{i \alpha}$ are the tree-level mass of $N_\alpha$ and Yukawa coupling, respectively. 

Charge conjugation transforms a particle to its antiparticle, and the charge conjugation operator $\mathsf{C}$ acting on spinors is defined by an operator that satisfies $\mathsf{C}^{-1} \gamma^\mu \mathsf{C} = \minus (\gamma^\mu)^\mathsf{T}$ where $\mathsf{T}$ denotes the transpose of a matrix. In the Dirac-Pauli representation of $\gamma^\mu$, it is given by $\mathsf{C} \coloneqq i \gamma^2 \gamma^0$, and the charge conjugate of a field $\psi$ is defined by $\psi^c \coloneqq \gamma^0 \mathsf{C} \psi^*$ where $\psi^* \coloneqq (\psi^\dag)^\mathsf{T}$. A fermionic field $N_\alpha$ is called Majorana, if it satisfies $N_\alpha^c = N_\alpha$, \textit{i.e.}, if the fermion generated by a Majorana field is its own antiparticle. Since charge conjugation changes the chirality of a field, we can write $N_\alpha = N_{R \alpha} + N_{R \alpha}^c$ where $N_{R \alpha} \coloneqq \mathsf{R} N_\alpha$.

Since a Majorana particle is its own antiparticle, there exist three different types of non-vanishing contraction between $N_\alpha$ and $\overline{N_\alpha}$:
\begin{align}
	\langle 0 | T \{N_\alpha (x) \overline{N_\alpha} (y)\} | 0 \rangle &= \int \frac{d^4 p}{(2 \pi)^4} \ e^{i p \cdot (x - y)} \frac{i}{\slashed{p} - m_{N_\alpha}}, \\
	\langle 0 | T \{N_\alpha (x) N_\alpha (y)\} | 0 \rangle &= \int \frac{d^4 p}{(2 \pi)^4} \ e^{i p \cdot (x - y)} \frac{i}{\slashed{p} - m_{N_\alpha}} \mathsf{C}, \\
	\langle 0 | T \{\overline{N_\alpha} (x) \overline{N_\alpha} (y)\} | 0 \rangle &= \int \frac{d^4 p}{(2 \pi)^4} \ e^{i p \cdot (x - y)} \mathsf{C} \frac{i}{\slashed{p} - m_{N_\alpha}}.
\end{align}
We call the first one \textit{Dirac-type}, and the others \textit{Majorana-type}. It is generally complicated to calculate an $S$-matrix element involving Majorana fields, not only because of the existence of several different types of propagators, but also because of the presence of $\mathsf{C}$ which requires careful tracking of spinor indices. In actual calculations, however, we can transform Majorana-type propagators into Dirac-type, and absorb $\mathsf{C}$ into existing fields in the $S$-matrix element so that its calculation is no more complicated than a diagram only with Dirac-type propagators. We will discuss it in detail in appendix \ref{sec:TrickMaj}.

\begin{figure}[t]
	\centering
	\subfloat[$i \slashed{p} \mathsf{R} (\Sigma_{0 R})_{\beta \alpha} (p^2)$]{
		\includegraphics[width = 40 mm]{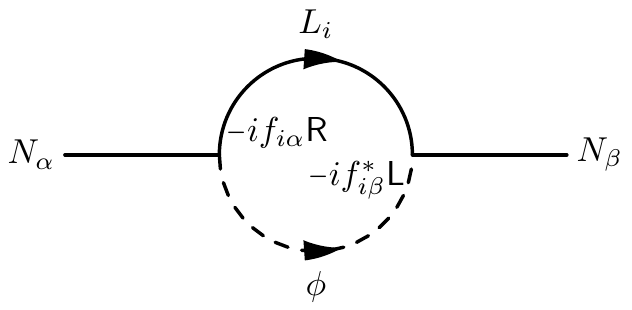}
	} \qquad \quad
	\subfloat[$i \slashed{p} \mathsf{L} (\Sigma_{0 L})_{\beta \alpha} (p^2)$]{
		\includegraphics[width = 40 mm]{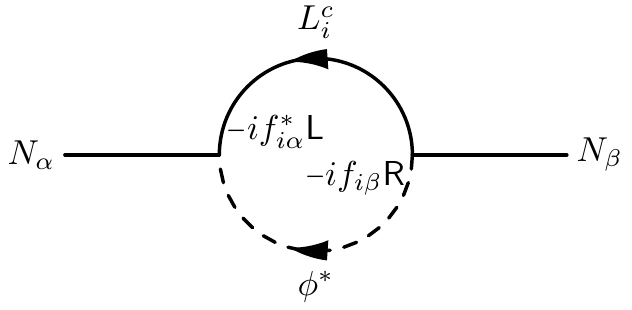}
	}
	\caption{One-loop diagrams which contribute to $i (\Sigma_0)_{\beta \alpha} (\slashed{p})$.}
	\label{fig:SE}
\end{figure}

The self-energy of a Majorana field is also different from that of a Dirac field since it has two contributions:
\begin{align}
	i \Sigma (\slashed{p}) = i \slashed{p} \big[ \mathsf{R} \Sigma_R (p^2) + \mathsf{L} \Sigma_L (p^2) \big],
\end{align}
which is a matrix defined over the flavor space of Majorana fields. The one-loop contributions to the self-energy of $N_\alpha$ are shown in figure \ref{fig:SE},
and they are given by
\begin{align}
	(\Sigma_{0 R})_{\beta \alpha} (p^2) = \sum_i \frac{f_{i \beta}^* f_{i \alpha}}{16 \pi^2} \bigg[ \minus \log{\bigg( \frac{p^2}{\Lambda^2} \bigg)} + i \pi \bigg], \qquad
	\Sigma_{0 L} (p^2) = (\Sigma_{0 R})^\mathsf{T} (p^2).
\end{align}
Here, we have assumed massless $L_i$ and $\phi$ for simplicity, and $\Lambda^2$ is defined by
\begin{align}
	\log{\Lambda^2} \coloneqq \frac{1}{\epsilon} - \gamma + \log{4 \pi} + 2,
\end{align}
where $1 / \epsilon$ is the regulator in dimensional regularization and $\gamma$ is the Euler-Mascheroni constant. Note that, in this paper, we will calculate each self-energy, \textit{i.e.}, each 1PI diagram, up to the one-loop order $\mathcal{O} (f^2 / 4 \pi)$. As mentioned in section \ref{sec:Intro}, the collective loop effect of mixing becomes non-perturbative when the unstable particle goes on-shell, and it should be carefully dealt with. Nevertheless, considering up to one loop for each 1PI diagram is still a good approximation as long as the Yukawa couplings are small, \textit{i.e.}, $|f| / 4 \pi \ll 1$.

These loop corrections cause mixing among different flavors of Majorana fields, and the main purpose of this paper is to discuss how to handle such loop-induced mixing and study its phenomenology.

\section{Renormalization for a single flavor}		\label{sec:Single}
In advance of discussing multiple flavors, we examine the mass and field-strength renormalization for a single flavor of a RH Majorana neutrino. For simplicity, only one LH lepton flavor will be considered here. The case of multiple flavors of RH Majorana neutrinos and LH leptons is a generalization of this simple case. Beginning from a bare Lagrangian, we will discuss renormalization of the self-energy and construct the renormalized Lagrangian. The renormalized resummed propagator as a function of the renormalized self-energy will be derived by a gemetric series, and the final form of the propagator will be obtained by expanding it around its pole and taking the leading contribution. The residue of the pole will also be explicitly calculated and we will see that it cannot be set to unity. \\

The bare Lagrangian involving the RH Majorana neutrino is given by
\begin{align}
	\mathcal{L} &= \frac{1}{2} \overline{N_0} i \slashed{\partial} N_0 - \frac{1}{2} m_0 \overline{N_0} \mathsf{R} N_0 - \frac{1}{2} m_0^* \overline{N_0} \mathsf{L} N_0
			- f_0 \overline{L_0} \widetilde{\phi}_0 \mathsf{R} N_0 - f_0^* \overline{N_0} \widetilde{\phi}_0^\dag \mathsf{L} L_0,
\end{align}
where $N_0$, $L_0$, and $\phi_0$ are the bare RH Majorana neutrino, LH lepton SU(2)-doublet, and scalar SU(2)-doublet, respectively. The renormalized fields and the corresponding field-strength renormalization factors are given by
\begin{align}
	N_{0 R} \eqqcolon Z_N^\frac{1}{2} N_R, \qquad
	N_{0 R}^c = Z_N^{\frac{1}{2} *} N_R^c, \qquad
	L_0 \eqqcolon Z_L^\frac{1}{2} L, \qquad
	\phi_0 \eqqcolon Z_\phi^\frac{1}{2} \phi,
\end{align}
and we also define the renormalized mass, mass renormalization factor, and renormalized Yukawa coupling as
\begin{align}
	m_0 \eqqcolon Z_M m_N, \qquad
 	f_0 Z_L^{\frac{1}{2} *} Z_\phi^{\frac{1}{2} *} Z_N^\frac{1}{2} \eqqcolon f.
\end{align}
Here, $m_N$ can be chosen to be a positive real number, for which $Z_M$ must be real. In addition, we do not need an additional vertex renormalization factor $Z_f$ for the Yukawa coupling: $f_0 Z_L^{\frac{1}{2} *} Z_\phi^{\frac{1}{2} *} Z_N^\frac{1}{2} = Z_f f$, since the vertex loop-correction is ultraviolet (UV) finite. Introducing a counterterm
\begin{align}
	Z_N^\frac{1}{2} \eqqcolon 1 + \frac{1}{2} \delta_N,
\end{align}
and defining
\begin{align}
	\delta m_N \coloneqq m_N Z_M Z_N - m_N
\end{align}
we write the bare Lagrangian in terms of renormalized fields and couplings as
\begin{align}
	\mathcal{L} &= \frac{1}{2} \overline{N_{0 R}} i \slashed{\partial} N_{0 R} + \frac{1}{2} \overline{N_{0 R}^c} i \slashed{\partial} N_{0 R}^c - \frac{1}{2} m_0 \overline{N_{0 R}^c} N_{0 R} - \frac{1}{2} m_0^* \overline{N_{0 R}} N_{0 R}^c
			- f_0 \overline{L_0} \widetilde{\phi}_0 N_{0 R} - f_0^* \overline{N_{0 R}} \widetilde{\phi}_0^\dag L_0 \nonumber \\
	&= \frac{1}{2} |Z_N| \overline{N_R} i \slashed{\partial} N_R + \frac{1}{2} |Z_N| \overline{N_R^c} i \slashed{\partial} N_R^c - \frac{1}{2} m_N Z_M Z_N \overline{N_R^c} N_R - \frac{1}{2} m_N Z_M^* Z_N^* \overline{N_R} N_R^c \nonumber \\
		&\qquad - f_0 Z_L^* Z_\phi^* Z_N \overline{L} \widetilde{\phi} N_R - f_0^* Z_L Z_\phi Z_N^* \overline{N_R} \widetilde{\phi}^\dag L \nonumber \\
	&= \frac{1}{2} \overline{N_R} i \slashed{\partial} N_R + \frac{1}{2} \overline{N_R^c} i \slashed{\partial} N_R^c - \frac{1}{2} m_N \overline{N_R^c} N_R - \frac{1}{2} m_N \overline{N_R} N_R^c
			- f \overline{L} \widetilde{\phi} N_R - f^* \overline{N_R} \widetilde{\phi}^\dag L \nonumber \\
		&\qquad + \frac{1}{4} (\delta_N^* + \delta_N + \cdots) \overline{N_R} i \slashed{\partial} N_R + \frac{1}{4} (\delta_N^* + \delta_N + \cdots) \overline{N_R^c} i \slashed{\partial} N_R^c
			 - \frac{1}{2} \delta m_N \overline{N_R^c} N_R - \frac{1}{2} \delta m_N^* \overline{N_R} N_R^c \nonumber \\
	&= \frac{1}{2} \overline{N} i \slashed{\partial} N - \frac{1}{2} m_N \overline{N} N - f \overline{L} \widetilde{\phi} \mathsf{R} N - f^* \overline{N} \widetilde{\phi}^\dag \mathsf{L} L \nonumber \\
		&\qquad + \frac{1}{4} (\delta_N^* + \delta_N + \cdots) \overline{N} i \slashed{\partial} N - \frac{1}{2} \delta m_N \overline{N} \mathsf{R} N - \frac{1}{2} \delta m_N^* \overline{N} \mathsf{L} N,
\end{align}
where the counterterms are explicitly written up to $\mathcal{O} (f^2 / 4 \pi)$.

The renormalized self-energy of the RH neutrino up to $\mathcal{O} (f^2 / 4 \pi)$ is now given by
\begin{align}
	\Sigma (\slashed{p}) \coloneqq \slashed{p} \bigg[ \Sigma_{0 R} (p^2) + \frac{1}{2} (\delta_N^* + \delta_N) \bigg] - \mathsf{R} \delta m_N - \mathsf{L} \delta m_N^*,
\end{align}
where for massless $L$ and $\phi$
\begin{align}
	\Sigma_{0 R} (p^2) \coloneqq \frac{|f|^2}{16 \pi^2} \bigg[ \minus \log \left( \frac{|p^2|}{\Lambda^2} \right) + i (\pi - \arg{[p^2]}) \bigg].
\end{align}
Note that we have allowed a complex value for $p^2$ in this expression by analytic continuation. In order to have a UV-finite self-energy $\Sigma (\slashed{p})$ for any $p$, $\delta m_N$ must be finite, which implies $\delta m_N$ can be regarded as a part of the renormalized complex mass. Since it is always possible to redefine the phase of $N_R$ to have a real mass, we may set $\delta m_N = 0$, \textit{i.e.}, $Z_M = Z_N^{-1}$, without loss of generality. Moreover, $\delta_N$ can be chosen as a real number since its imaginary part has no role in renormalization. Hence, the bare Lagrangian can be rewritten as 
\begin{empheq}[box=\fbox]{align}
	\mathcal{L} &= \frac{1}{2} \overline{N_0} i \slashed{\partial} N_0 - \frac{1}{2} m_0 \overline{N_0} N_0 - f_0 \overline{L_0} \widetilde{\phi}_0 \mathsf{R} N_0 - f_0^* \overline{N_0} \widetilde{\phi}_0^\dag \mathsf{L} L_0 \nonumber \\
		&= \frac{1}{2} \overline{N} i \slashed{\partial} N - \frac{1}{2} m_N \overline{N} N - f \overline{L} \widetilde{\phi} \mathsf{R} N - f^* \overline{N} \widetilde{\phi}^\dag \mathsf{L} L
		+ \frac{1}{4} (2 \delta_N + \delta_N^2) \overline{N} i \slashed{\partial} N,
	\label{eq:LagSingle}
\end{empheq}
and the renormalized self-energy is now given by
\begin{align}
	\boxed{\Sigma (\slashed{p}) = \slashed{p} \Sigma_R (p^2),}
\end{align}
where
\begin{align}
	\boxed{\Sigma_R (p^2) = \Sigma_{0 R} (p^2) + \delta_N.}
\end{align}
Here, $\delta_N$ is a real number, which implies the imaginary part of $\Sigma_R (p^2)$ should be identical to that of $\Sigma_{0 R} (p^2)$. This is a constraint on the renormalization condition for the self-energy, as is well-known.

Even though there exist several different types of propagators for Majorana fields, it is always possible to use only the Dirac-type propagator, which will be explained in appendix \ref{sec:TrickMaj}. Hence, we consider only a Dirac-type propagator in the following discussion. The resummed propagator can therefore be obtained by a geometric series as follows:
\begin{align}
	i \Delta (\slashed{p}) &= \frac{i}{\slashed{p} - m_N} + \frac{i}{\slashed{p} - m_N} [i \Sigma (\slashed{p})] \frac{i}{\slashed{p} - m_N}
		+ \frac{i}{\slashed{p} - m_N} [i \Sigma (\slashed{p})] \frac{i}{\slashed{p} - m_N} [i \Sigma (\slashed{p})] \frac{i}{\slashed{p} - m_N} + \cdots \nonumber \\
	&= \sum_{n = 0}^\infty \frac{i}{\slashed{p} - m_N} \left\{ [i \Sigma (\slashed{p})] \frac{i}{\slashed{p} - m_N} \right\}^n \nonumber \\
	&= i \big\{ \slashed{p} - m_N + \Sigma (\slashed{p}) \big\}^{-1}
	= i \big\{ [1 + \Sigma_R (p^2)] \slashed{p} - m_N \big\}^{-1}.
\end{align}
To invert the expression of $\Delta^{-1} (\slashed{p})$, we write
\begin{align}
	\Delta (\slashed{p}) = \mathsf{R} \Delta_{RR} (p^2) + \mathsf{R} \slashed{p} \Delta_{RL} (p^2) + \mathsf{L} \slashed{p} \Delta_{LR} (p^2) + \mathsf{L} \Delta_{LL} (p^2).
\end{align}
Since
\begin{align}
	1 = \Delta^{-1} (\slashed{p}) \Delta (\slashed{p}) &= \big\{ [1 + \Sigma_R (p^2)] \slashed{p} - m_N \big\} \mathsf{R} [\Delta_{RR} (p^2) + \slashed{p} \Delta_{RL} (p^2)] \nonumber \\
		&\qquad + \big\{ [1 + \Sigma_R (p^2)] \slashed{p} - m_N \big\} \mathsf{L} [\Delta_{LL} (p^2) + \slashed{p} \Delta_{LR} (p^2)],
\end{align}
we can write
\begin{align}
	0 &= \mathsf{L} [\Delta^{-1} (\slashed{p}) \Delta (\slashed{p})] \mathsf{R}
		= \mathsf{L} \slashed{p} \big\{ [1 + \Sigma_R (p^2)] \Delta_{RR} (p^2) - m_N \Delta_{LR} (p^2) \big\}, \\
	\mathsf{R} &= \mathsf{R} [\Delta^{-1} (\slashed{p}) \Delta (\slashed{p})] \mathsf{R}
		= \mathsf{R} \big\{ \minus m_N \Delta_{RR} (p^2) + [1 + \Sigma_R (p^2)] p^2 \Delta_{LR} (p^2) \big\}.
\end{align}
Hence,
\begin{align}
	\Delta_{RR} (p^2) &= \frac{m_N}{[1 + \Sigma_R (p^2)]^2 p^2 - m_N^2}
		= \frac{m_N [1 + \Sigma_R (p^2)]^{-2}}{p^2 - m_N^2 [1 + \Sigma_R (p^2)]^{-2}}, \\
	\Delta_{LR} (p^2) &= \frac{1}{m_N} [1 + \Sigma_R (p^2)] \Delta_{RR} (p^2).
\end{align}
In addition, from $0 = \mathsf{R} (\Delta^{-1} \Delta) \mathsf{L}$ and $\mathsf{L} = \mathsf{L} (\Delta^{-1} \Delta) \mathsf{L}$, we also obtain
\begin{align}
	\Delta_{LL} (p^2) &= \frac{m_N}{[1 + \Sigma_R (p^2)]^2 p^2 - m_N^2}
		= \frac{m_N [1 + \Sigma_R (p^2)]^{-2}}{p^2 - m_N^2 [1 + \Sigma_R (p^2)]^{-2}}, \\
	\Delta_{RL} (p^2) &= \frac{1}{m_N} [1 + \Sigma_R (p^2)] \Delta_{LL} (p^2).
\end{align}
Defining
\begin{align}
	P (p^2) \coloneqq m_N [1 + \Sigma_R (p^2)]^{-1},
\end{align}
we write
\begin{align}
	i \Delta_{RR} (p^2) &= i \Delta_{LL} (p^2) = \frac{P (p^2)}{m_N} \frac{i P (p^2)}{p^2 - P^2 (p^2)},
		\label{eq:PropAB1Single} \\
	i \slashed{p} \Delta_{LR} (p^2) &= i \slashed{p} \Delta_{RL} (p^2) = \frac{P (p^2)}{m_N} \frac{i \slashed{p}}{p^2 - P^2 (p^2)}.
		\label{eq:PropAB2Single}
\end{align}
The resummed propagator is therefore written as
\begin{align}
	i \Delta (\slashed{p}) = \frac{P (p^2)}{m_N} \frac{i [\slashed{p} + P (p^2)]}{p^2 - P^2 (p^2)}.
	\label{eq:PropSingle}
\end{align}
This is usually not the final form of the propagator in practical application, and the next step is to expand it around its pole and take the leading contribution. To be consistent with the Breit-Wigner resonance pattern, the physical pole $p_{\widehat{N}}^2$ of the propagator must be in the form of
\begin{align}
	p_{\widehat{N}}^2 = m_{\widehat{N}}^2 - i m_{\widehat{N}} \Gamma_{\widehat{N}},
\end{align}
where $m_{\widehat{N}}$ and $\Gamma_{\widehat{N}}$ are the pole mass and total decay width of $N$. The complex mass $p_{\widehat{N}}$ is the solution of the equation
\begin{align}
	p = P (p^2) = m_N [1 + \Sigma_R (p^2)]^{-1}.
\end{align}
Up to $\mathcal{O} (f^2 / 4 \pi)$, we can write $P (p^2) = m_N [1 - \Sigma_R (p^2)]$ and $p_{\widehat{N}} = m_{\widehat{N}} - i \Gamma_{\widehat{N}} / 2$, and thus
\begin{align}
	\boxed{\text{Re} [\Sigma_R (p_{\widehat{N}}^2)] = \frac{m_{\widehat{N}}}{m_N} - 1, \qquad
	\text{Im} [\Sigma_R (p_{\widehat{N}}^2)] = \frac{\Gamma_{\widehat{N}}}{2 m_N}.}
	\label{eq:SEtoObsSingle}
\end{align}
Since
\begin{align}
	\lim_{p^2 \to p_{\widehat{N}}^2} \frac{p^2 - P^2 (p^2)}{p^2 - p_{\widehat{N}}^2}
	= 1 - \frac{dP^2}{dp^2} (p_{\widehat{N}}^2),
\end{align}
the residues of the pole are written as
\begin{align}
	\lim_{p^2 \to p_{\widehat{N}}^2} (p^2 - p_{\widehat{N}}^2) \Delta_{RR} (p^2) &= \lim_{p^2 \to p_{\widehat{N}}^2} (p^2 - p_{\widehat{N}}^2) \Delta_{LL} (p^2)
		= R_{\widehat{N}} p_{\widehat{N}}, \\
	\lim_{p^2 \to p_{\widehat{N}}^2} (p^2 - p_{\widehat{N}}^2) \Delta_{LR} (p^2) &= \lim_{p^2 \to p_{\widehat{N}}^2} (p^2 - p_{\widehat{N}}^2) \Delta_{RL} (p^2)
		= R_{\widehat{N}}
\end{align}
where we have defined
\begin{align}
	R_{\widehat{N}} \coloneqq \frac{p_{\widehat{N}}}{m_N} \bigg[ 1 - \frac{dP^2}{dp^2} (p_{\widehat{N}}^2) \bigg]^{-1}.
\end{align}
Using
\begin{align}
	\frac{d\Sigma_{0 R}}{dp^2} = \minus \frac{|f|^2}{16 \pi^2} \frac{1}{p^2},
\end{align}
we can write
\begin{align}
	R_{\widehat{N}} &= \frac{p_{\widehat{N}}}{m_N} \bigg[ 1 - 2 m_N^2 \frac{d\Sigma_R}{dp^2} (p_{\widehat{N}}^2) \bigg]
	= \frac{p_{\widehat{N}}}{m_N} \bigg[ 1 - 2 m_N^2 \frac{d\Sigma_{0 R}}{dp^2} (p_{\widehat{N}}^2) \bigg]
	= \frac{p_{\widehat{N}}}{m_N} \bigg( 1 + \frac{|f|^2}{8 \pi^2} \frac{m_N^2}{p_{\widehat{N}}^2} \bigg).
\end{align}
Hence, up to $\mathcal{O} (f^2 / 4 \pi)$
\begin{align}
	\boxed{R_{\widehat{N}} = \frac{p_{\widehat{N}}}{m_N} \bigg( 1 + \frac{|f|^2}{8 \pi^2} \bigg),}
	\label{eq:ResSingle}
\end{align}
which shows that the residue is \textit{not} unity because $p_{\widehat{N}}$ is complex while all the other factors are real. Furthermore, $Z_M = Z_N^{-1}$ implies
\begin{align}
	\boxed{\delta_M = \minus \delta_N = \text{Re} [\Sigma_{0 R} (p_{\widehat{N}}^2)] + 1 - \frac{m_{\widehat{N}}}{m_N},}
\end{align}
where equation \ref{eq:SEtoObsSingle} has been used. For on-shell renormalization, we must choose $\text{Re} [\Sigma_R (p_{\widehat{N}}^2)] = 0$ to have $m_N = m_{\widehat{N}}$. The renormalized resummed propagator given by equations \ref{eq:PropAB1Single}-\ref{eq:PropSingle} can finally be rewritten as
\begin{empheq}[box=\fbox]{align}
	i \Delta_{RR} (p^2) &= i \Delta_{LL} (p^2)
		= R_{\widehat{N}} \frac{i p_{\widehat{N}}}{p^2 - p_{\widehat{N}}^2} + \cdots, \\
	i \slashed{p} \Delta_{LR} (p^2) &= i \slashed{p} \Delta_{RL} (p^2)
		= R_{\widehat{N}} \frac{i \slashed{p}}{p^2 - p_{\widehat{N}}^2} + \cdots, \\
	i \Delta (\slashed{p}) &= i \mathsf{R} \Delta_{RR} (p^2) + i \mathsf{R} \slashed{p} \Delta_{RL} (p^2) + i \mathsf{L} \slashed{p} \Delta_{LR} (p^2) + i \mathsf{L} \Delta_{LL} (p^2) \nonumber \\
		& = R_{\widehat{N}} \frac{i (\slashed{p} + p_{\widehat{N}})}{p^2 - p_{\widehat{N}}^2} + \cdots.
\end{empheq}

\section{Renormalization for multiple flavors}	\label{sec:Multiple}
Now let us generalize the renormalization procedure to the case of multiple flavors, where the self-energy and resummed propagator are matrices defined over the flavor space of RH Majorana neutrinos. We will follow the same steps as in the case of a single flavor: the renormalized Lagrangian will be constructed from a given bare Lagrangian, and meanwhile the self-energy matrix of RH Majorana neutrinos will be renormalized. The resummed propagator as a function of the self-energy will be obtained by a gemetric series at a matrix level. \\

The bare Lagrangian for multiple flavors is written as
\begin{align}
	\mathcal{L} &= \frac{1}{2} \sum_\alpha \overline{N_{0 \alpha}} i \slashed{\partial} N_{0 \alpha} - \frac{1}{2} \sum_{\alpha, \beta} (M_0)_{\beta \alpha} \overline{N_{0 \beta}} \mathsf{R} N_{0 \alpha} - \frac{1}{2} \sum_{\alpha, \beta} (M_0)_{\beta \alpha}^* \overline{N_{0 \beta}} \mathsf{L} N_{0 \alpha} \nonumber \\
		&\qquad - \sum_{i, \alpha} (f_0)_{i \alpha} \overline{L_{0 i}} \widetilde{\phi}_0 \mathsf{R} N_{0 \alpha} - \sum_{i, \alpha} (f_0)_{i \alpha}^* \overline{N_{0 \alpha}} \widetilde{\phi}_0^\dag \mathsf{L} L_{0 i}.
\end{align}
The renormalized fields and field-strength renormalization factors are given by
\begin{alignat}{2}
	&N_{0 R \beta} \eqqcolon \sum_\alpha (Z_N^\frac{1}{2})_{\beta \alpha} N_{R \alpha}, \qquad
	&&N_{0 R \beta}^c = \sum_\alpha (Z_N^\frac{1}{2})_{\beta \alpha}^* N_{R \alpha}^c, \\
	&L_{0 j} \eqqcolon \sum_i (Z_L^\frac{1}{2})_{ji} L_i, \qquad
	&&\phi_0 \eqqcolon Z_\phi^\frac{1}{2} \phi.
\end{alignat}
In addition, the renormalized mass, mass renormalization factor, and renormalized Yukawa couplings are written as
\begin{align}
	(M_0)_{\beta \alpha} \eqqcolon (Z_M^{\frac{1}{2} \mathsf{T}} M_N Z_M^\frac{1}{2})_{\beta \alpha}, \qquad
 	Z_\phi^{\frac{1}{2} *} (Z_L^{\frac{1}{2} \dag} f_0 Z_N^\frac{1}{2})_{i \alpha} \eqqcolon f_{i \alpha}.
\end{align}
Without loss of generality, we may choose real diagonal $M_N$, \textit{i.e.}, $(M_N)_{\beta \alpha} \eqqcolon m_{N_\alpha} \delta_{\beta \alpha}$ where $m_{N_\alpha}$ is a real positive number. Moreover, as in the single-flavor case, a vertex counterterm is not needed since the vertex-loop correction is UV finite. Introducing a field-strength counterterm matrix
\begin{align}
	Z_N^\frac{1}{2} \eqqcolon 1 + \frac{1}{2} \delta_N,
\end{align}
and defining
\begin{align}
	\delta M_N \coloneqq Z_N^{\frac{1}{2} \mathsf{T}} Z_M^{\frac{1}{2} \mathsf{T}} M_N Z_M^\frac{1}{2} Z_N^\frac{1}{2} - M_N,
\end{align}
we rewrite the bare Lagrangian as
\begin{align}
	\mathcal{L} &= \frac{1}{2} \sum_\alpha \overline{N_{0 R \alpha}} i \slashed{\partial} N_{0 R \alpha} + \frac{1}{2} \sum_\alpha \overline{N_{0 R \alpha}^c} i \slashed{\partial} N_{0 R \alpha}^c - \frac{1}{2} \sum_{\alpha, \beta} (M_0)_{\beta \alpha} \overline{N_{0 R \beta}^c} N_{0 R \alpha} - \frac{1}{2} \sum_{\alpha, \beta} (M_0)_{\beta \alpha}^* \overline{N_{0 R \beta}} N_{0 R \alpha}^c \nonumber \\
		&\qquad - \sum_{i, \alpha} (f_0)_{i \alpha} \overline{L_{0 i}} \widetilde{\phi}_0 N_{0 R \alpha} - \sum_{i, \alpha} (f_0)_{i \alpha}^* \overline{N_{0 R \alpha}} \widetilde{\phi}_0^\dag L_{0 i} \nonumber \\
	&= \frac{1}{2} \sum_{\alpha, \beta} (Z_N^{\frac{1}{2} \dag} Z_N^\frac{1}{2})_{\beta \alpha} \overline{N_{R \beta}} i \slashed{\partial} N_{R \alpha} + \frac{1}{2} \sum_{\alpha, \beta} (Z_N^{\frac{1}{2} \dag} Z_N^\frac{1}{2})_{\beta \alpha}^* \overline{N_{R \beta}^c} i \slashed{\partial} N_{R \alpha}^c \nonumber \\
		&\qquad - \frac{1}{2} \sum_{\alpha, \beta} (Z_N^{\frac{1}{2} \mathsf{T}} Z_M^{\frac{1}{2} \mathsf{T}} M_N Z_M^\frac{1}{2} Z_N^\frac{1}{2})_{\beta \alpha} \overline{N_{R \beta}^c} N_{R \alpha} - \frac{1}{2} \sum_{\alpha, \beta} (Z_N^{\frac{1}{2} \mathsf{T}} Z_M^{\frac{1}{2} \mathsf{T}} M_N Z_M^\frac{1}{2} Z_N^\frac{1}{2})_{\beta \alpha}^* \overline{N_{R \beta}} N_{R \alpha}^c \nonumber \\
		&\qquad - \sum_{i, \alpha} Z_\phi^{\frac{1}{2} *} (Z_L^{\frac{1}{2} \dag} f_0 Z_N^\frac{1}{2})_{i \alpha} \overline{L_i} \widetilde{\phi} N_{R \alpha} - \sum_{i, \alpha} Z_\phi^\frac{1}{2} (Z_L^{\frac{1}{2} \dag} f_0 Z_N^{\frac{1}{2}})_{i \alpha}^* \overline{N_{R \alpha}} \widetilde{\phi}^\dag L_i \nonumber \\
	&= \frac{1}{2} \sum_\alpha \overline{N_{R \alpha}} i \slashed{\partial} N_{R \alpha} + \frac{1}{2} \sum_\alpha \overline{N_{R \alpha}^c} i \slashed{\partial} N_{R \alpha}^c
			- \frac{1}{2} \sum_\alpha m_{N_\alpha} \overline{N_{R \alpha}^c} N_{R \alpha} - \frac{1}{2} \sum_\alpha m_{N_\alpha} \overline{N_{R \alpha}} N_{R \alpha}^c \nonumber \\
			&\qquad \qquad - \sum_{i, \alpha} f_{i \alpha} \overline{L_i} \widetilde{\phi} N_{R \alpha} - \sum_{i, \alpha} f_{i \alpha}^* \overline{N_{R \alpha}} \widetilde{\phi}^\dag L_i \nonumber \\
		&\qquad + \frac{1}{4} \sum_{\alpha, \beta} (\delta_N^\dag + \delta_N + \cdots)_{\beta \alpha} \overline{N_{R \beta}} i \slashed{\partial} N_{R \alpha} + \frac{1}{4} \sum_{\alpha, \beta} (\delta_N^\dag + \delta_N + \cdots)_{\beta \alpha}^* \overline{N_{R \beta}^c} i \slashed{\partial} N_{R \alpha}^c \nonumber \\
			&\qquad \qquad - \frac{1}{2} \sum_{\alpha, \beta} (\delta M_N)_{\beta \alpha} \overline{N_{R \beta}^c} N_{R \alpha}
				- \frac{1}{2} \sum_{\alpha, \beta} (\delta M_N)_{\beta \alpha}^* \overline{N_{R \beta}} N_{R \alpha}^c \nonumber \\
	&= \frac{1}{2} \sum_\alpha \overline{N_\alpha} i \slashed{\partial} N_\alpha - \frac{1}{2} \sum_\alpha m_{N_\alpha} \overline{N_\alpha} N_\alpha
			- \sum_{i, \alpha} f_{i \alpha} \overline{L_i} \widetilde{\phi} \mathsf{R} N_\alpha - \sum_{i, \alpha} f_{i \alpha}^* \overline{N_\alpha} \widetilde{\phi}^\dag \mathsf{L} L_i \nonumber \\
		&\qquad + \frac{1}{4} \sum_{\alpha, \beta} (\delta_N^\dag + \delta_N + \cdots)_{\beta \alpha} \overline{N_\beta} i \slashed{\partial} \mathsf{R} N_\alpha + \frac{1}{4} \sum_{\alpha, \beta} (\delta_N^\dag + \delta_N + \cdots)_{\beta \alpha}^* \overline{N_\beta} i \slashed{\partial} \mathsf{L} N_\alpha \nonumber \\
			&\qquad \qquad - \frac{1}{2} \sum_{\alpha, \beta} (\delta M_N)_{\beta \alpha} \overline{N_\beta} \mathsf{R} N_\alpha
				- \frac{1}{2} \sum_{\alpha, \beta} (\delta M_N)_{\beta \alpha}^* \overline{N_\beta} \mathsf{L} N_\alpha,
\end{align}
where the counterterms are written up to $\mathcal{O} (f^2 / 4 \pi)$ for simplicity. 

The renormalized self-energy up to $\mathcal{O} (f^2 / 4 \pi)$ is given by
\begin{align}
	\Sigma (\slashed{p}) &= \slashed{p} \mathsf{R} \bigg[ \Sigma_{0 R} (p^2) + \frac{1}{2} (\delta_N^\dag + \delta_N) \bigg] + \slashed{p} \mathsf{L} \bigg[ \Sigma_{0 R} (p^2) + \frac{1}{2} (\delta_N^\dag + \delta_N) \bigg]^\mathsf{T}
	+ \mathsf{R} \delta M_N + \mathsf{L} \delta M_N^*
\end{align}
where for massless $L_i$ and $\phi$
\begin{align}
	(\Sigma_{0 R})_{\beta \alpha} (p^2) \coloneqq \sum_i \frac{f_{i \beta}^* f_{i \alpha}}{16 \pi^2} \bigg[ \minus \log{\left( \frac{|p^2|}{\Lambda^2} \right)} + i (\pi - \arg{[p^2]}) \bigg].
\end{align}
Here, we have allowed a complex value for $p^2$ again by means of analytic continuation. In order to have UV-finite $\Sigma (\slashed{p})$ for any $p$, $\delta M_N$ must be finite, which in turn implies $\delta M_N$ may be regarded as a part of the renormalized mass of $N_\alpha$. Since $M_N$ is the renormalized mass matrix of RH neutrinos by choice, we may set $\delta M_N = 0$, \textit{i.e.}, $Z_M = Z_N^{-1}$, without loss of generality. Furthermore, we may choose a Hermitian matrix for $\delta_N$, because its skew-Hermitian part has no role in renormalizing $\Sigma_{0 R} (p^2)$. Note that an arbitrary matrix $X$ can be written as $X = X_H + X_S$ where $X_H \coloneqq (X + X^\dag) / 2$ is Hermitian and $X_S \coloneqq (X - X^\dag) / 2$ is skew-Hermitian. The renormalized Lagrangian and counterterms are now written as
\begin{empheq}[box=\fbox]{align}
	\mathcal{L} &= \frac{1}{2} \sum_\alpha \overline{N_\alpha} i \slashed{\partial} N_\alpha - \frac{1}{2} \sum_\alpha m_{N_\alpha} \overline{N_\alpha} N_\alpha
			- \sum_{i, \alpha} f_{i \alpha} \overline{L_i} \widetilde{\phi} \mathsf{R} N_\alpha - \sum_{i, \alpha} f_{i \alpha}^* \overline{N_\alpha} \widetilde{\phi}^\dag \mathsf{L} L_i \nonumber \\
		&\qquad + \frac{1}{4} \sum_{\alpha, \beta} (2 \delta_N + \delta_N^2)_{\beta \alpha} \overline{N_\beta} i \slashed{\partial} N_\alpha,
	\label{eq:LagMultiple}
\end{empheq}
and the renormalized self-energy matrix up to $\mathcal{O} (f^2 / 4 \pi)$ is
\begin{align}
	\boxed{\Sigma (\slashed{p}) = \slashed{p} \mathsf{R} \Sigma_R (p^2) + \slashed{p} \mathsf{L} \Sigma_R^\mathsf{T} (p^2),}
	\label{eq:SE}
\end{align}
where we have defined
\begin{align}
	\boxed{\Sigma_R (p^2) \coloneqq \Sigma_{0 R} (p^2) + \delta_N.}
	\label{eq:SER}
\end{align}
Here, $\delta_N$ can be chosen as a Hermitian matrix without loss of generality as mentioned above, and thus the skew-Hermitian parts of $\Sigma_R (p^2)$ and $\Sigma_{0 R} (p^2)$ must be identical. This is a constraint on the renormalization condition for the self-energy, and $\delta_N$ will be accordingly chosen in section \ref{sec:Diag}.

\begin{figure}[t]
	\centering
	\subfloat[$i \slashed{p} \mathsf{R} (\Sigma_{0 R})_{\beta \alpha} (p^2)$]{
		\includegraphics[width = 40 mm]{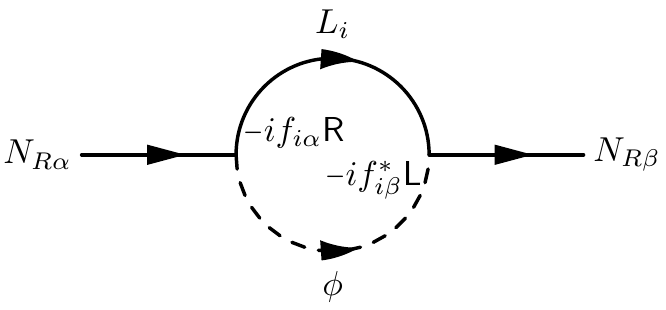}
		\label{fig:SERD}
	} \qquad \quad
	\subfloat[$i \slashed{p} \mathsf{L} (\Sigma_{0 L})_{\beta \alpha} (p^2)$]{
		\includegraphics[width = 40 mm]{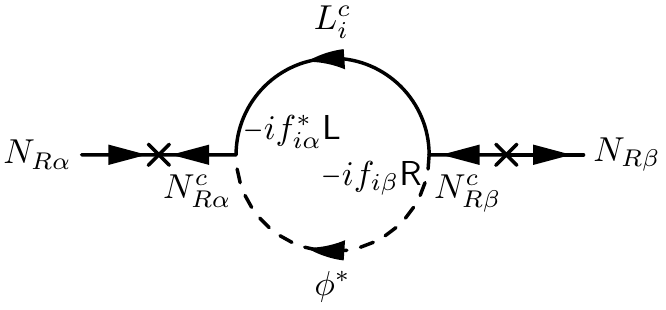}
		\label{fig:SELD}
	}
	\caption{One-loop diagrams which contribute to $i (\Delta_{RR})_{\beta \alpha} (p^2)$.}
	\label{fig:SED}
	\subfloat[$i \slashed{p} \mathsf{R} (\Sigma_{0 R})_{\beta \alpha} (p^2)$]{
		\includegraphics[width = 40 mm]{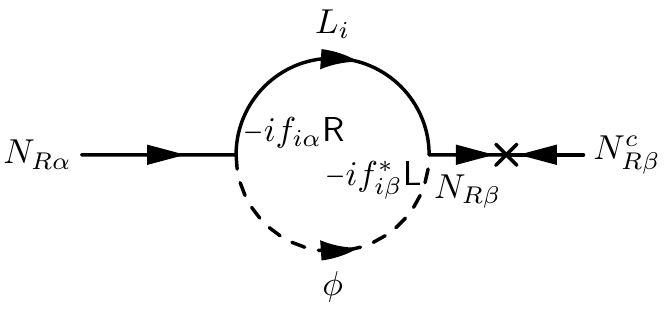}
		\label{fig:SERM}
	} \qquad \quad
	\subfloat[$i \slashed{p} \mathsf{L} (\Sigma_{0 L})_{\beta \alpha} (p^2)$]{
		\includegraphics[width = 40 mm]{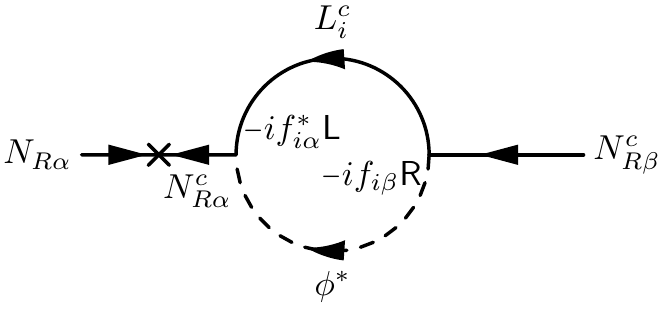}
		\label{fig:SELM}
	}
	\caption{One-loop diagrams which contribute to $i (\Delta_{LR})_{\beta \alpha} (p^2)$.}
	\label{fig:SEM}
\end{figure}

As in the single-flavor case, it is sufficient to consider only Dirac-type propagators in actual calculations. Hence, the resummed propagator can be obtained by the geometric series at a matrix level as follows:
\begin{align}
	i \Delta (\slashed{p}) &= i (\slashed{p} - M_N)^{-1} + i (\slashed{p} - M_N)^{-1} [i \Sigma (\slashed{p})] i (\slashed{p} - M_N)^{-1} \nonumber \\
		&\qquad + i (\slashed{p} - M_N)^{-1} [i \Sigma (\slashed{p})] i (\slashed{p} - M_N)^{-1} [i \Sigma (\slashed{p})] i (\slashed{p} - M_N)^{-1} + \cdots \nonumber \\
	&= \sum_{n = 0}^\infty i (\slashed{p} - M_N)^{-1} \big\{ [i \Sigma (\slashed{p})] i (\slashed{p} - M_N)^{-1} \big\}^n,
\end{align}
from which we deduce
\begin{align}
	\minus i \Delta^{-1} (\slashed{p}) &= \minus i [\slashed{p} - M_N + \Sigma (\slashed{p})]
	= \minus i \big\{ [1 + \mathsf{L} \Sigma_R (p^2) + \mathsf{R} \Sigma_R^\mathsf{T} (p^2)] \slashed{p} - M_N \big\}.
\end{align}
To invert $\Delta^{-1} (\slashed{p})$, we write $\Delta (\slashed{p})$ as
\begin{align}
	\Delta (\slashed{p}) = \mathsf{R} \Delta_{RR} (p^2) + \mathsf{R} \slashed{p} \Delta_{RL} (p^2) + \mathsf{L} \slashed{p} \Delta_{LR} (p^2) + \mathsf{L} \Delta_{LL} (p^2).
\end{align}
Note that both $\Sigma_R (p^2)$ and $\Sigma_L (p^2) = \Sigma_R^\mathsf{T} (p^2)$ contribute to each chiral component $\Delta_{AB} (p^2)~(A, B = L, R)$. To illustrate it, some examples of one-loop diagrams are given in figures \ref{fig:SED} and \ref{fig:SEM}.
Using this chiral decomposition, we obtain
\begin{align}
	1 = \Delta^{-1} (\slashed{p}) \Delta (\slashed{p})
	&= \big\{ [1 + \mathsf{L} \Sigma_R (p^2)] \slashed{p} - M_N \big\} \mathsf{R} [\Delta_{RR} (p^2) + \slashed{p} \Delta_{RL} (p^2)] \nonumber \\
		&\qquad + \big\{ [1 + \mathsf{R} \Sigma_R^\mathsf{T} (p^2)] \slashed{p} - M_N \big\} \mathsf{L} [\Delta_{LL} (p^2) + \slashed{p} \Delta_{LR} (p^2)].
\end{align}
From
\begin{align}
	0 &= \mathsf{L} [\Delta^{-1} (\slashed{p}) \Delta (\slashed{p})] \mathsf{R}
		= \mathsf{L} \slashed{p} \big\{ [1 + \Sigma_R (p^2)] \Delta_{RR} (p^2) - M_N \Delta_{LR} (p^2) \big\}, \\
	\mathsf{R} &= \mathsf{R} [\Delta^{-1} (\slashed{p}) \Delta (\slashed{p})] \mathsf{R}
		= \mathsf{R} \big\{ \minus M_N \Delta_{RR} (p^2) + [1 + \Sigma_R^\mathsf{T} (p^2)] p^2 \Delta_{LR} (p^2) \big\},
\end{align}
it is easy to find
\begin{align}
	\Delta_{RR} (p^2) &= \big\{ [1 + \Sigma_R^\mathsf{T} (p^2)] M_N^{-1} [1 + \Sigma_R (p^2)] p^2 - M_N \big\}^{-1},
		\label{eq:PropRRMultiple} \\
	\Delta_{LR} (p^2) &= M_N^{-1} [1 + \Sigma_R (p^2)] \Delta_{RR} (p^2).
		\label{eq:PropLRMultiple}
\end{align}
Similarly, $0 = \mathsf{R} [\Delta^{-1} (\slashed{p}) \Delta (\slashed{p})] \mathsf{L}$ and $0 = \mathsf{L} [\Delta^{-1} (\slashed{p}) \Delta (\slashed{p})] \mathsf{L}$ imply
\begin{align}
	\Delta_{LL} (p^2) &= \big\{ [1 + \Sigma_R (p^2)] M_N^{-1} [1 + \Sigma_R^\mathsf{T} (p^2)] p^2 - M_N \big\}^{-1},
		\label{eq:PropLLMultiple} \\
	\Delta_{RL} (p^2) &= M_N^{-1} [1 + \Sigma_R^\mathsf{T} (p^2)] \Delta_{LL} (p^2).
		\label{eq:PropRLMultiple}
\end{align}

\section{Diagonalization of the propagator matrix}		\label{sec:Diag}
In this section, the resummed propagator will be diagonalized and the associated mixing matrices will be obtained. Expanding each component of the diagonalized propagator around its pole, we will also obtain the final form of the propagator useful for practical applications. The residues and effective Yukawa couplings will be found as well, and the self-energy satisfying the constraints on renormalization conditions will also be presented. \\

First we define a pole of $\Delta_{AB} (p^2) $ as a solution of
\begin{align}
	\text{det} [\Delta_{AB}^{-1} (p^2)] = 0,
\end{align}
where the determinant is taken over the flavors. This definition implies that a similarity transformation would be involved in diagonalizing the propagator, and thus the corresponding mixing matrix is non-unitary in general.

Since the chiral components of the propagator are all different, it may look unclear whether their poles are identical. Here we prove that $\Delta_{RR}$, $\Delta_{LR}$, $\Delta_{RL}$, and $\Delta_{LL}$ indeed have identical poles. Since equation \ref{eq:PropLRMultiple} implies
\begin{align}
	\text{det} [\Delta_{LR} (p^2)] = \text{det} \big[ M_N^{-1} \{1 + \Sigma_R (p^2)\} \big] \text{det} [\Delta_{RR} (p^2)]
\end{align}
and $\text{det} \big[ \{1 + \Sigma_R\}^{-1} M_N \big] \neq 0$, the solutions of $\text{det} [\Delta_{LR}^{-1}] = 0$ is the same as those of $\text{det} [\Delta_{RR}^{-1}] = 0$. Hence, $\Delta_{LR}$ and $\Delta_{RR}$ have identical poles. We can similarly show that $\Delta_{RL}$ and $\Delta_{LL}$ have identical poles. Furthermore, we may rewrite equations \ref{eq:PropRRMultiple} and \ref{eq:PropLLMultiple} as
\begin{align}
	M_N^{-\frac{1}{2}} \Delta_{RR}^{-1} (p^2) M_N^{-\frac{1}{2}} &= M_N^{-\frac{1}{2}} [1 + \Sigma_R^\mathsf{T} (p^2)] M_N^{-\frac{1}{2}} M_N^{-\frac{1}{2}} [1 + \Sigma_R (p^2)] M_N^{-\frac{1}{2}} p^2 - 1, \\
	M_N^{-\frac{1}{2}} \Delta_{LL}^{-1} (p^2) M_N^{-\frac{1}{2}} &= M_N^{-\frac{1}{2}} [1 + \Sigma_R (p^2)] M_N^{-\frac{1}{2}} M_N^{-\frac{1}{2}} [1 + \Sigma_R^\mathsf{T} (p^2)] M_N^{-\frac{1}{2}} p^2 - 1.
\end{align}
By Sylvester's theorem which says $\text{det} [1 - XY] = \text{det} [1 - YX]$ for arbitraty matrices $X$ and $Y$, we deduce $\text{det} [\Delta_{RR}^{-1}] = \text{det} [\Delta_{LL}^{-1}]$. Hence, $\Delta_{RR}$ and $\Delta_{LL}$ have identical poles. \textit{q.e.d.} \\

Now let us discuss how to diagonalize the propagator. Defining
\begin{align}
	\boxed{A (p^2) \coloneqq 1 + \Sigma_R (p^2),}
	\label{eq:A}
\end{align}
we can compactly write
\begin{alignat}{2}
	&\Delta_{RR} (p^2) = (A^\mathsf{T} M_N^{-1} A \ p^2 - M_N)^{-1}, \qquad
	&&\Delta_{LR} (p^2) = M_N^{-1} A \Delta_{RR}, \\
	&\Delta_{LL} (p^2) = (A M_N^{-1} A^\mathsf{T} p^2 - M_N)^{-1}, \qquad
	&&\Delta_{RL} (p^2) = M_N^{-1} A^\mathsf{T} \Delta_{LL}.
\end{alignat}
We first consider the diagonalization of $\Delta_{RR}$. Introducing another shorthand notation
\begin{align}
	\boxed{B (p^2) \coloneqq M_N^\frac{1}{2} A^{-1} (p^2) M_N^\frac{1}{2},}
	\label{eq:B}
\end{align}
we write
\begin{align}
	\Delta_{RR} = (A^{-1} M_N^\frac{1}{2}) (p^2 - B^\mathsf{T} B)^{-1} (A^{-1} M_N^\frac{1}{2})^\mathsf{T}.
\end{align}
Since $B^\mathsf{T} B$ is a symmetric matrix, there exists an orthogonal matrix $O_L$ such that
\begin{align}
	\boxed{P^2 (p^2) \coloneqq O_L^\mathsf{T} (p^2) B^\mathsf{T} (p^2) B (p^2) O_L (p^2)}
\end{align}
is diagonal. Furthermore, $P (p^2)$ can be chosen as a diagonal matrix with $\text{Re} [P_{\widehat{\alpha}} (p^2)] > 0$, where $P_{\widehat{\alpha}} \coloneqq P_{\widehat{\alpha} \widehat{\alpha}}$. Note that we use Latin characters with a hat for the indices of the diagonalized propagator and the associated field. Now $\Delta_{RR}$ can be written as
\begin{align}
	\Delta_{RR} = (M_N^{-\frac{1}{2}} B O_L P^{-1} M_N^\frac{1}{2}) (M_N^{-1} P^2) (p^2 - P^2)^{-1} (M_N^{-\frac{1}{2}} B O_L P^{-1} M_N^\frac{1}{2})^\mathsf{T}
\end{align}
where we have used the fact that $M_N$ and $P$ are diagonal matrices, \textit{e.g.}, $M_N P = P M_N$. We also define
\begin{align}
	\boxed{O_R (p^2) \coloneqq B (p^2) O_L (p^2) P^{-1} (p^2),}
\end{align}
\textit{i.e.},
\begin{align}
	\boxed{P (p^2) = O_R^\mathsf{T} (p^2) B (p^2) O_L (p^2) = O_L^\mathsf{T} (p^2) B^\mathsf{T} (p^2) O_R (p^2),}
\end{align}
which gives
\begin{align}
	\Delta_{RR} = (M_N^{-\frac{1}{2}} O_R M_N^\frac{1}{2}) (M_N^{-1} P^2) (p^2 - P^2)^{-1} (M_N^{-\frac{1}{2}} O_R M_N^\frac{1}{2})^\mathsf{T}.
\end{align}
This matrix $O_R$ is an orthogonal matrix since
\begin{align}
	O_R^\mathsf{T} O_R &= (B O_L P^{-1})^\mathsf{T} (B O_L P^{-1})
	= P^{-1} O_L^\mathsf{T} B^\mathsf{T} B O_L P^{-1}
	= P^{-1} P^2 P^{-1}
	= 1.
\end{align}
Moreover, we have
\begin{align}
	O_R^\mathsf{T} B B^\mathsf{T} O_R &= (B O_L P^{-1})^\mathsf{T} (B B^\mathsf{T}) (B O_L P^{-1})
	= P^{-1} (O_L^\mathsf{T} B^\mathsf{T} B O_L) (O_L^\mathsf{T} B^\mathsf{T} B O_L) P^{-1}
	= P^2,
\end{align}
which is needed to diagonalize $\Delta_{LL}$:
\begin{align}
	\Delta_{LL} &= [(A^\mathsf{T})^{-1} M_N^\frac{1}{2}] (p^2 - B B^\mathsf{T})^{-1} [(A^\mathsf{T})^{-1} M_N^\frac{1}{2}]^\mathsf{T} \nonumber \\
	&= [(A^\mathsf{T})^{-1} M_N^\frac{1}{2}] O_R (p^2 - P^2)^{-1} O_R^\mathsf{T} [(A^\mathsf{T})^{-1} M_N^\frac{1}{2}]^\mathsf{T} \nonumber \\
	&= (M_N^{-\frac{1}{2}} O_L M_N^\frac{1}{2}) (M_N^{-1} P^2) (p^2 - P^2)^{-1} (M_N^{-\frac{1}{2}} O_L M_N^\frac{1}{2})^\mathsf{T}.
\end{align}
Defining
\begin{align}
	\boxed{C_R (p^2) \coloneqq M_N^{-\frac{1}{2}} O_R (p^2) M_N^\frac{1}{2}, \qquad
	C_L (p^2) \coloneqq M_N^{-\frac{1}{2}} O_L (p^2) M_N^\frac{1}{2},}
\end{align}
we can finally write
\begin{align}
	C_R^{-1} \Delta_{RR} (C_R^\mathsf{T})^{-1}
	= C_L^{-1} \Delta_{LL} (C_L^\mathsf{T})^{-1}
	= (M_N^{-1} P^2) (p^2 - P^2)^{-1}.
\end{align}
In addition,
\begin{align}
	C_L^{-1} \Delta_{LR} (C_R^\mathsf{T})^{-1} &= C_L^{-1} (M_N^{-1} A \Delta_{RR}) (C_R^\mathsf{T})^{-1}
	= (C_L^{-1} M_N^{-1} A C_R) [C_R^{-1} \Delta_{RR} (C_R^\mathsf{T})^{-1}] \nonumber \\
	&= (C_L^{-1} M_N^{-1} A C_R) [M_N^{-1} P^2 (p^2 - P^2)^{-1}] \nonumber \\
	&= [M_N^{-\frac{1}{2}} (O_R^\mathsf{T} B O_L)^{-1} M_N^{-\frac{1}{2}} P^2] (p^2 - P^2)^{-1} \nonumber \\
	&= (M_N^{-1} P) (p^2 - P^2)^{-1}.
\end{align}
Similarly,
\begin{align}
	C_R^{-1} \Delta_{RL} (C_L^\mathsf{T})^{-1} 
	= (M_N^{-1} P) (p^2 - P^2)^{-1}.
\end{align}
Hence, $C_R (p^2)$ and $C_L (p^2)$ are the mixing matrices which diagonalize the resummed propagator matrix. 

Note that we have not defined $O_R$ as an orthogonal matrix that diagonalizes $B B^\mathsf{T}$, since this definition leaves an unwanted uncertainty in $O_R$ relative to $O_L$. For example, if a matrix $O$ is defined as an orthogonal matrix which diagonalizes a symmetric matrix $Q$ by $O^\mathsf{T} Q O$, we have a freedom in choosing the sign of each column of $O$ when constructing it out of the corresponding eigenvector. If both $O_L$ and $O_R$ are defined as such, we might have unwanted signs in the elements of the diagonalized forms of $\Delta_{RL}$ and $\Delta_{LR}$ depending on our choice, since their diagonalization involves both $O_L$ and $O_R$. Even though we may allow some freedom in $O_L$, the freedom in $O_R$ should be eliminated for the proper diagonalization of $\Delta_{AB}$ for all $A, B$.

Let us denote the diagonalized propagator and the fields corresponding to its diagonal components by $i \widehat{\Delta} (\slashed{p})$ and $\widehat{N}_{\widehat{\alpha}}$, respectively. The diagonalized propagator can be written as
\begin{align}
	\widehat{\Delta} (\slashed{p}) = \mathsf{R} \widehat{\Delta}_{RR} (p^2) + \mathsf{R} \slashed{p} \widehat{\Delta}_{RL} (p^2) + \mathsf{L} \slashed{p} \widehat{\Delta}_{LR} (p^2) + \mathsf{L} \widehat{\Delta}_{LL} (p^2),
\end{align}
where each chiral component satisfies
\begin{align}
	\boxed{\Delta_{AB} (p^2) = C_A (p^2) \widehat{\Delta}_{AB} (p^2) C_B^\mathsf{T} (p^2).}
	\label{eq:PropAB}
\end{align}
Then,
\begin{align}
	\boxed{\widehat{\Delta}_{\widehat{\alpha}} (p^2) = \frac{P_{\widehat{\alpha}} (p^2)}{m_{N_\alpha}} \frac{\slashed{p} + P_{\widehat{\alpha}} (p^2)}{p^2 - P_{\widehat{\alpha}}^2 (p^2)},}
	\label{eq:PropMD}
\end{align}
where $ \delta_{\widehat{\beta} \widehat{\alpha}} \widehat{\Delta}_{\widehat{\alpha}} \coloneqq \widehat{\Delta}_{\widehat{\beta} \widehat{\alpha}}$, and
\begin{empheq}[box=\fbox]{align}
	(\widehat{\Delta}_{RR})_{\widehat{\alpha}} (p^2) &= (\widehat{\Delta}_{LL})_{\widehat{\alpha}} (p^2) = \frac{P_{\widehat{\alpha}} (p^2)}{m_{N_\alpha}} \frac{P_{\widehat{\alpha}} (p^2)}{p^2 - P_{\widehat{\alpha}}^2 (p^2)},
		\label{eq:PropMDAB1} \\
	\slashed{p} (\widehat{\Delta}_{LR})_{\widehat{\alpha}} (p^2) &= \slashed{p} (\widehat{\Delta}_{RL})_{\widehat{\alpha}} (p^2) = \frac{P_{\widehat{\alpha}} (p^2)}{m_{N_\alpha}} \frac{\slashed{p}}{p^2 - P_{\widehat{\alpha}}^2 (p^2)}.
		\label{eq:PropMDAB2}
\end{empheq}
It should be emphasized that this diagonalization is \textit{exact} as long as
\begin{align}
	\boxed{\Sigma_L (p^2) = \Sigma_R^\mathsf{T} (p^2)}
	\label{eq:SELR}
\end{align}
is satisfied. This relationship is correct up to $\mathcal{O} (f^2 / 4 \pi)$ where the self-energy is given by equation \ref{eq:SE}, but it is unclear whether it still holds up to a higher order in perturbation. Our working precision is at most $\mathcal{O} (f^2 / 4 \pi)$, and the diagonalization above can thus be considered to be exact. If equation \ref{eq:SELR} is satisfied up to the infinite order in perturbation, then the diagonalization discussed here is exact up to the infinite order as well. \\

To obtain the expression of $P (p^2)$, we show the followings:
\begin{align}
	O_L^\mathsf{T} (p^2) O_R (p^2) = 1 + \mathcal{O} (f^2 / 4 \pi),
\end{align}
which is true whether the complex mixing angle of $O_A (p^2)$ is small or large. This is trivially satisfied when $\Delta m_N \gg \Gamma_N$, since $O_A (p^2) \sim 1 + \mathcal{O} (f^2 / 4 \pi)$ in such a case, as we will see in section \ref{sec:Example}. On the other hand, when $\Delta m_N \lesssim \Gamma_N$, we will also show that $O_A (p^2)$ is a large-mixing matrix, and thus it is unclear whether $O_L^\mathsf{T} O_R$ is indeed close to the identity. In that case, we can write $m_{N_\beta} = m_{N_\alpha} + \mathcal{O} (f^2 / 4 \pi)$, and thus
\begin{align}
	B^{-1} (p^2) = M_N^{-\frac{1}{2}} A M_N^{-\frac{1}{2}}
	= \frac{1}{m_{N_\alpha}} [1 + \mathcal{O} (f^2 / 4 \pi)].
\end{align}
We therefore have
\begin{align}
	P^{-1} (p^2) = O_L^\mathsf{T} B^{-1} O_R
	= \frac{1}{m_{N_\alpha}} O_L^\mathsf{T} O_R + \mathcal{O} (f / 4 \pi),
\end{align}
which implies that $O_L^\mathsf{T} O_R$ is positive diagonal up to the leading order in perturbation, since $P^{-1} (p^2)$ is a diagonal matrix with a positive real part by choice. This in turn implies $O_L^\mathsf{T} O_R = 1$ up to the leading order. To explicitly see that for two flavors, we parametrize $O_A (p^2)$ by
\begin{align}
	O_A (p^2) = \left( \begin{array}{cc} \cos{z_A} & \sin{z_A} \\
		\minus \sin{z_A} & \cos{z_A} \end{array} \right),
\end{align}
where $z_A (p^2)$ is a complex-valued angle. Then, we can write
\begin{align}
	O_L^\mathsf{T} (p^2) O_R (p^2) = \left( \begin{array}{cc} \cos{z_L} & \minus \sin{z_L} \\
			\sin{z_L} & \cos{z_L} \end{array} \right)
		\left( \begin{array}{cc} \cos{z_R} & \sin{z_R} \\
			\minus \sin{z_R} & \cos{z_R} \end{array} \right)
	= \left( \begin{array}{cc} \cos{(z_R - z_L)} & \sin{(z_R - z_L)} \\
			\minus \sin{(z_R - z_L)} & \cos{(z_R - z_L)} \end{array} \right).
\end{align}
The off-diagonal components vanish up to the leading order, only if $z_R = z_L$ up to the same precision. Hence, we conclude $O_L^\mathsf{T} O_R = 1 + \mathcal{O} (f^2 / 4 \pi)$. As a consequence, we also have
\begin{align}
	\boxed{C_L^\mathsf{T} (p^2) C_R (p^2) = 1 + \mathcal{O} (f^2 / 4 \pi),}
	\label{eq:CLTCR}
\end{align}
which will be useful for several purposes. \textit{q.e.d.} \\

Now let us derive the expression of $P_{\widehat{\alpha}} (p^2)$. It is written as
\begin{align}
	P (p^2) = O_R^\mathsf{T} B O_L
	= M_N^\frac{1}{2} C_R^{-1} A^{-1} (C_L^\mathsf{T})^{-1} M_N^\frac{1}{2}
	= M_N \big[ C_L^\mathsf{T} (1 + \Sigma_R) C_R \big]^{-1}.
\end{align}
Defining the diagonal self-energy matrix $\widehat{\Sigma}_R (p^2)$ by
\begin{align}
	\boxed{1 + \widehat{\Sigma}_R (p^2)
	\coloneqq C_L^\mathsf{T} (p^2) A (p^2) C_R (p^2)
	= C_L^\mathsf{T} (p^2) [1 + \Sigma_R (p^2)] C_R (p^2),}
\end{align}
we write
\begin{align}
	\boxed{P (p^2) = M_N [1 + \widehat{\Sigma}_R (p^2)]^{-1}.}
\end{align}
Note that $\widehat{\Sigma}_R (p^2) \sim \mathcal{O} (f^2 / 4 \pi)$ due to equation \ref{eq:CLTCR}. We also introduce a shorthand notation for each diagonal component of $\widehat{\Sigma}_R (p^2)$: $\delta_{\widehat{\beta} \widehat{\alpha}} (\widehat{\Sigma}_R)_{\widehat{\alpha}} (p^2) \coloneqq (\widehat{\Sigma}_R)_{\widehat{\beta} \widehat{\alpha}} (p^2)$.

To be consistent with the Breit-Wigner resonance pattern of the scattering cross section, the complex pole of the propagator should be in the form of
\begin{align}
	p_{\widehat{N}_{\widehat{\alpha}}}^2 = m_{\widehat{N}_{\widehat{\alpha}}}^2 - i m_{\widehat{N}_{\widehat{\alpha}}} \Gamma_{\widehat{N}_{\widehat{\alpha}}}.
\end{align}
where $m_{\widehat{N}_{\widehat{\alpha}}}$ and $\Gamma_{\widehat{N}_{\widehat{\alpha}}}$ are the pole mass and total decay width of $\widehat{N}_{\widehat{\alpha}}$. The complex mass $p_{\widehat{N}_{\widehat{\alpha}}}$ is a solution of the equation
\begin{align}
	p = P_{\widehat{\alpha}} (p^2)
	= m_{N_\alpha} [1 + (\widehat{\Sigma}_R)_{\widehat{\alpha}} (p^2)]^{-1}
	= m_{N_\alpha} \big\{ C_L^\mathsf{T} (p^2) [1 + \Sigma_R (p^2)] C_R (p^2) \big\}_{\widehat{\alpha} \widehat{\alpha}}^{-1}.
\end{align}
Up to $\mathcal{O} (f^2 / 4 \pi)$, we can write $P_{\widehat{\alpha}} (p^2) = m_{N_\alpha} [1 - (\widehat{\Sigma}_R)_{\widehat{\alpha}} (p^2)]$ and $p_{\widehat{N}_{\widehat{\alpha}}} = m_{\widehat{N}_{\widehat{\alpha}}} - i \Gamma_{\widehat{N}_{\widehat{\alpha}}} / 2$, and thus
\begin{empheq}[box=\fbox]{align}
	\text{Re} [(\widehat{\Sigma}_R)_{\widehat{\alpha}} (p_{\widehat{N}_{\widehat{\alpha}}}^2)] &= \text{Re} \bigg[ c_{\widehat{\alpha}} + \sum_{\beta, \gamma} d^{\widehat{\alpha}}_{\beta \gamma} (\Sigma_R)_{\beta \gamma} (p_{\widehat{N}_{\widehat{\alpha}}}^2) \bigg]
		= \frac{m_{\widehat{N}_{\widehat{\alpha}}}}{m_{N_\alpha}} - 1, \\
	\text{Im} [(\widehat{\Sigma}_R)_{\widehat{\alpha}} (p_{\widehat{N}_{\widehat{\alpha}}}^2)] &= \text{Im} \bigg[ c_{\widehat{\alpha}} + \sum_{\beta, \gamma} d^{\widehat{\alpha}}_{\beta \gamma} (\Sigma_R)_{\beta \gamma} (p_{\widehat{N}_{\widehat{\alpha}}}^2) \bigg]
		= \frac{\Gamma_{\widehat{N}_{\widehat{\alpha}}}}{2 m_{N_\alpha}},
		\label{eq:ImSE}
\end{empheq}
where we have defined
\begin{align}
	C_R^{\widehat{N}_{\widehat{\alpha}}} \coloneqq C_R (p_{\widehat{N}_{\widehat{\alpha}}}^2), \qquad
	c_{\widehat{\alpha}} \coloneqq \sum_\beta (C_L^{\widehat{N}_{\widehat{\alpha}}})_{\beta \widehat{\alpha}} (C_R^{\widehat{N}_{\widehat{\alpha}}})_{\beta \widehat{\alpha}} - 1, \qquad
	d^{\widehat{\alpha}}_{\beta \gamma} \coloneqq (C_L^{\widehat{N}_{\widehat{\alpha}}})_{\beta \widehat{\alpha}} (C_R^{\widehat{N}_{\widehat{\alpha}}})_{\gamma \widehat{\alpha}}.
\end{align}
Note that equation \ref{eq:CLTCR} also implies $c_{\widehat{\alpha}} \lesssim \mathcal{O} (f^2 / 4 \pi)$. When the mass difference between RH neutrinos is small, it turns out that $c_{\widehat{\alpha}}$ and $d^{\widehat{\alpha}}_{\beta \gamma}$ themselves are complicated functions of $(\Sigma_R)_{\beta \gamma} (p_{\widehat{N}_{\widehat{\alpha}}}^2)$, and thus, in such a case, $(\widehat{\Sigma}_R)_{\widehat{\alpha}} (p_{\widehat{N}_{\widehat{\alpha}}}^2)$ is much different from $(\Sigma_R)_{\alpha \alpha} (p_{\widehat{N}_{\widehat{\alpha}}}^2)$. Some examples will be given in section \ref{sec:Example}.

The final form of the propagator is obtained by expanding it around the complex pole and taking the leading part. Since
\begin{align}
	\lim_{p^2 \to p_{\widehat{N}_{\widehat{\alpha}}}^2} \frac{p^2 - P_{\widehat{\alpha}}^2 (p^2)}{p^2 - p_{\widehat{N}_{\widehat{\alpha}}}^2} = 1 - \frac{dP_{\widehat{\alpha}}^2}{dp^2} (p_{\widehat{N}_{\widehat{\alpha}}}^2),
\end{align}
the residues of the pole are written as
\begin{align}
	\lim_{p^2 \to p_{\widehat{N}_{\widehat{\alpha}}}^2} (p^2 - p_{\widehat{N}_{\widehat{\alpha}}}^2) (\widehat{\Delta}_{RR})_{\widehat{\alpha} \widehat{\alpha}} (p^2)
		&= \lim_{p^2 \to p_{\widehat{N}_{\widehat{\alpha}}}^2} (p^2 - p_{\widehat{N}_{\widehat{\alpha}}}^2) (\widehat{\Delta}_{LL})_{\widehat{\alpha} \widehat{\alpha}} (p^2)
		= R_{\widehat{N}_{\widehat{\alpha}}} p_{\widehat{N}_{\widehat{\alpha}}}, \\
	\lim_{p^2 \to p_{\widehat{N}_{\widehat{\alpha}}}^2} (p^2 - p_{\widehat{N}_{\widehat{\alpha}}}^2) (\widehat{\Delta}_{LR})_{\widehat{\alpha} \widehat{\alpha}} (p^2)
		&= \lim_{p^2 \to p_{\widehat{N}_{\widehat{\alpha}}}^2} (p^2 - p_{\widehat{N}_{\widehat{\alpha}}}^2) (\widehat{\Delta}_{RL})_{\widehat{\alpha} \widehat{\alpha}} (p^2)
		= R_{\widehat{N}_{\widehat{\alpha}}}
\end{align}
where
\begin{align}
	R_{\widehat{N}_{\widehat{\alpha}}} \coloneqq \frac{p_{\widehat{N}_{\widehat{\alpha}}}}{m_{N_\alpha}} \bigg[ 1 - \frac{dP_{\widehat{\alpha}}^2}{dp^2} (p_{\widehat{N}_{\widehat{\alpha}}}^2) \bigg]^{-1}.
\end{align}
Using
\begin{align}
	\frac{d(\Sigma_R)_{\beta \gamma}}{dp^2} = \minus \sum_i \frac{f_{i \beta}^* f_{i \gamma}}{16 \pi^2} \frac{1}{p^2},
\end{align}
we can write up to $\mathcal{O} (f^2 / 4 \pi)$
\begin{align}
	R_{\widehat{N}_{\widehat{\alpha}}} = \frac{p_{\widehat{N}_{\widehat{\alpha}}}}{m_{N_\alpha}} \bigg[ 1 - 2 m_{N_\alpha}^2 \frac{d(\widehat{\Sigma}_R)_\alpha}{dp^2} (p_{\widehat{N}_{\widehat{\alpha}}}^2) \bigg]
	= \frac{p_{\widehat{N}_{\widehat{\alpha}}}}{m_{N_\alpha}} \bigg[ 1 + \frac{1}{8 \pi^2} \sum_i (f^* C_L^{\widehat{N}_{\widehat{\alpha}}})_{i \widehat{\alpha}} (f C_R^{\widehat{N}_{\widehat{\alpha}}})_{i \widehat{\alpha}} \bigg].
\end{align}
Defining the effective Yukawa couplings by the loop corrections to the field-strength as
\begin{align}
	\boxed{\widehat{f}_{i \widehat{\alpha}} \coloneqq (f C_R^{\widehat{N}_{\widehat{\alpha}}})_{i \widehat{\alpha}}, \qquad
	\widehat{f}^c_{i \widehat{\alpha}} \coloneqq (f^* C_L^{\widehat{N}_{\widehat{\alpha}}})_{i \widehat{\alpha}},}
	\label{eq:feff}
\end{align}
we write
\begin{align}
	\boxed{R_{\widehat{N}_{\widehat{\alpha}}} = \frac{p_{\widehat{N}_{\widehat{\alpha}}}}{m_{N_\alpha}} \bigg( 1 + \frac{1}{8 \pi^2} \sum_i \widehat{f}^c_{i \widehat{\alpha}} \widehat{f}_{i \widehat{\alpha}} \bigg).}
\end{align}
Similarly to the single-flavor case, the residue $R_{\widehat{N}_{\widehat{\alpha}}}$ \textit{cannot} be set to unity in general whichever renormalization condition is chosen.

As we have discussed in the derivation of equation \ref{eq:SER}, the counterterm $\delta_N$ can be chosen as a Hermitian matrix without loss of generality. For example, we may choose
\begin{align}
	(\delta_N)_{\beta \alpha} = \sum_i \frac{f_{i \beta}^* f_{i \alpha}}{16 \pi^2} \log{\bigg( \frac{|p_{\widehat{N}_{\widehat{\beta}}} p_{\widehat{N}_{\widehat{\alpha}}}|}{\Lambda^2} \bigg)},
	\label{eq:deltaN}
\end{align}
which implies
\begin{align}
	(\Sigma_R)_{\beta \alpha} (p^2) = \sum_i \frac{f_{i \beta}^* f_{i \alpha}}{16 \pi^2} \bigg[ \minus \log{\bigg( \frac{|p^2|}{|p_{\widehat{N}_{\widehat{\beta}}} p_{\widehat{N}_{\widehat{\alpha}}}|} \bigg)} + i (\pi - \arg{[p^2]}) \bigg].
\end{align}
Moreover, we have chosen the mass renormalization factor by $Z_M = Z_N^{-1}$ to write the Lagrangian as equation \ref{eq:LagMultiple}. In the case of a single flavor, the choice of $\delta_N$ given by equation \ref{eq:deltaN} corresponds to an on-shell renormalization scheme in the sense of $m_{\widehat{N}} = m_N$ since it makes $\Sigma_R (p^2)$ have only an imaginary part. In the case of multiple flavors, however, equation \ref{eq:deltaN} does not result in $m_{\widehat{N}_{\widehat{\alpha}}} = m_{N_\alpha}$ in general, since $(\widehat{\Sigma}_R)_{\widehat{\alpha}} (p^2)$ is not guaranteed to have only a real part because of the non-zero off-diagonal components in $\Sigma_R (p^2)$. In fact, for a small mass difference, \textit{i.e.}, $\Delta m_N \lesssim \Gamma_N$, the condition $m_{\widehat{N}_{\widehat{\alpha}}} = m_{N_\alpha}$ has no specific meaning, since $\widehat{N}_{\widehat{\alpha}}$ associated with $m_{\widehat{N}_{\widehat{\alpha}}}$ is totally different from $N_\alpha$ associated with $m_{N_\alpha}$. In other words, the on-shell renormalization scheme, which is supposed to impose $\widehat{N}_{\widehat{\alpha}} = N_\alpha$ by choosing $m_{\widehat{N}_{\widehat{\alpha}}} = m_{N_\alpha}$, cannot work for $\Delta m_N \lesssim \Gamma_N$. We will discuss this with more details in section \ref{sec:Quasi}. In actual calculations up to $\mathcal{O} (f^2 / 4 \pi)$ with real-valued $p^2$, we can use the expression
\begin{align}
	\boxed{(\Sigma_R)_{\beta \alpha} (p^2) = \sum_i \frac{f_{i \beta}^* f_{i \alpha}}{16 \pi^2} \bigg[ \minus \log{\bigg( \frac{p^2}{m_{N_\beta} m_{N_\alpha}} \bigg)} + i \pi \bigg]}
	\label{eq:SEexp}
\end{align}
since $\log{|p_{\widehat{N}_{\widehat{\alpha}}}|} = \log{m_{N_\alpha}} + \mathcal{O} (f^2 / 4 \pi)$.

The components of the renormalized diagonal resummed propagator given by equations \ref{eq:PropMD}-\ref{eq:PropMDAB2} can now be rewritten as 
\begin{empheq}[box=\fbox]{align}
	i (\widehat{\Delta}_{RR})_{\widehat{\alpha}} (p^2) &= i (\widehat{\Delta}_{LL})_{\widehat{\alpha}} (p^2)
		= R_{\widehat{N}_{\widehat{\alpha}}} \frac{i p_{\widehat{N}_{\widehat{\alpha}}}}{p^2 - p_{\widehat{N}_{\widehat{\alpha}}}^2} + \cdots, \\
	i \slashed{p} (\widehat{\Delta}_{LR})_{\widehat{\alpha}} (p^2) &= i \slashed{p} (\widehat{\Delta}_{RL})_{\widehat{\alpha}} (p^2)
		= R_{\widehat{N}_{\widehat{\alpha}}} \frac{i \slashed{p}}{p^2 - p_{\widehat{N}_{\widehat{\alpha}}}^2} + \cdots, \\
	i \widehat{\Delta}_{\widehat{\alpha}} (\slashed{p}) &= \big[ i \mathsf{R} \widehat{\Delta}_{RR} (p^2) + i \mathsf{R} \slashed{p} \widehat{\Delta}_{RL} (p^2) + i \mathsf{L} \slashed{p} \widehat{\Delta}_{LR} (p^2) + i \mathsf{L} \widehat{\Delta}_{LL} (p^2) \big]_{\alpha \alpha} \nonumber \\
		&= \frac{i R_{\widehat{N}_{\widehat{\alpha}}} }{\slashed{p} - p_{\widehat{N}_{\widehat{\alpha}}}} + \cdots.
\end{empheq}
Moreover, we may write equation \ref{eq:PropAB} as
\begin{empheq}[box=\fbox]{align}
	i (\Delta_{RR})_{\beta \alpha} (p^2) &= \sum_{\widehat{\gamma}} (C_R^{\widehat{N}_{\widehat{\gamma}}})_{\beta {\widehat{\gamma}}} \frac{i R_{\widehat{N}_{\widehat{\gamma}}} p_{\widehat{N}_{\widehat{\gamma}}}}{p^2 - p_{\widehat{N}_{\widehat{\gamma}}}^2} (C_R^{\widehat{N}_{\widehat{\gamma}}})_{\alpha {\widehat{\gamma}}} + \cdots,
		\label{eq:PropMDRRExp} \\
	i (\Delta_{LL})_{\beta \alpha} (p^2) &= \sum_{\widehat{\gamma}} (C_L^{\widehat{N}_{\widehat{\gamma}}})_{\beta {\widehat{\gamma}}} \frac{i R_{\widehat{N}_{\widehat{\gamma}}} p_{\widehat{N}_{\widehat{\gamma}}}}{p^2 - p_{\widehat{N}_{\widehat{\gamma}}}^2} (C_L^{\widehat{N}_{\widehat{\gamma}}})_{\alpha {\widehat{\gamma}}} + \cdots,
		\label{eq:PropMDLLExp} \\
	i (\Delta_{RL})_{\beta \alpha} (p^2) &= \sum_{\widehat{\gamma}} (C_R^{\widehat{N}_{\widehat{\gamma}}})_{\beta {\widehat{\gamma}}} \frac{i R_{\widehat{N}_{\widehat{\gamma}}}}{p^2 - p_{\widehat{N}_{\widehat{\gamma}}}^2} (C_L^{\widehat{N}_{\widehat{\gamma}}})_{\alpha {\widehat{\gamma}}} + \cdots,
		\label{eq:PropMDRLExp} \\
	i (\Delta_{LR})_{\beta \alpha} (p^2) &= \sum_{\widehat{\gamma}} (C_L^{\widehat{N}_{\widehat{\gamma}}})_{\beta {\widehat{\gamma}}} \frac{i R_{\widehat{N}_{\widehat{\gamma}}}}{p^2 - p_{\widehat{N}_{\widehat{\gamma}}}^2} (C_R^{\widehat{N}_{\widehat{\gamma}}})_{\alpha {\widehat{\gamma}}} + \cdots,
		\label{eq:PropMDLRExp}
\end{empheq}
which provide a simpler analysis of mixing than equation \ref{eq:PropAB} does. The residue $R_{\widehat{N}_{\widehat{\gamma}}}$ can be set to unity in calculations up to $\mathcal{O} (f^2 / 4 \pi)$.

Using the effective Yukawa couplings defined by equation \ref{eq:feff} and ignoring the vertex-loop correction, we can rewrite, for example, the scattering amplitude of $L_i^k \phi^k \to L_j^l \phi^l$ as
\begin{align}
	i \mathcal{M} (L_i^k \phi^k \to L_j^l \phi^l) &= \sum_{\alpha, \beta} \overline{u_{L_j^l}} (\mathbf{p}_{L_j^l}) (\minus i f_{j \beta} \mathsf{R}) [i \Delta_{\beta \alpha} (\slashed{p})] (\minus i f_{i \alpha}^* \mathsf{L}) u_{L_i^k} (\mathbf{p}_{L_i^k}) \nonumber \\
	&= \sum_{\widehat{\alpha}} \overline{u_{L_j^l}} (\mathbf{p}_{L_j^l}) (\minus i \widehat{f}_{j \widehat{\alpha}} \mathsf{R}) \frac{i R_{\widehat{N}_{\widehat{\alpha}}} \slashed{p}}{p^2 - p_{\widehat{N}_{\widehat{\alpha}}}^2} (\minus i \widehat{f}^c_{i \widehat{\alpha}} \mathsf{L}) u_{L_i^k} (\mathbf{p}_{L_i^k}) + \cdots.
\end{align}
Here, $k$, $l$ are SU(2) indices without the Einstein summation convention. As in this example, we may sometimes replace the Yukawa couplings and non-diagonal propagator with the effective Yukawa couplings and diagonalized propagator for practical purposes. However, such a prescription is not allowed, if the replacement implies that the degree of freedom corresponding to the component of the diagonalized propagator should be regarded as an external state of a physical process. We now discuss what it means.

\section{Generation of quasiparticles}		\label{sec:Quasi}
In this section, we identify the degree of freedom associated with $\widehat{\Delta}_{\widehat{\alpha}}$. We will see that it should be interpreted as a quasiparticle which loses Majorana nature. \\

The correlation function corresponding to the non-diagonal propagator is
\begin{align}
	\int \frac{d^4 p}{(2 \pi)^4} e^{-i p \cdot (x - y)} i \Delta_{\beta \alpha} (\slashed{p})
	= \langle \Omega | N_\beta (x) \overline{N_\alpha} (y) | \Omega \rangle, \ (x^0 > y^0).
\end{align}
Note that it should be already time-ordered since $\Delta_{\beta \alpha}$ is defined with a specific direction of energy transfer. Let us first consider $\Delta_{RR}$, whose component is given by
\begin{align}
	i (\Delta_{RR})_{\beta \alpha} (p^2) &= \sum_{\widehat{\gamma}} (C_R^{\widehat{N}_{\widehat{\gamma}}})_{\beta {\widehat{\gamma}}} \frac{i R_{\widehat{N}_{\widehat{\gamma}}} p_{\widehat{N}_{\widehat{\gamma}}}}{p^2 - p_{\widehat{N}_{\widehat{\gamma}}}^2} (C_R^{\widehat{N}_{\widehat{\gamma}}})_{\alpha {\widehat{\gamma}}} + \cdots.
\end{align}
The associated chiral component of the two-point function is
\begin{align}
	\int \frac{d^4 p}{(2 \pi)^4} e^{-i p \cdot (x - y)} \mathsf{R} i (\Delta_{RR})_{\beta \alpha} (p^2)  \mathsf{R}
	= \langle \Omega | \mathsf{R} N_\beta (x) \overline{N_\alpha} (y) \mathsf{R} | \Omega \rangle, \ (x^0 > y^0),
\end{align}
and thus the correlation function corresponding to the component of the diagonalized propagator is written as
\begin{align}
	\int \frac{d^4 p}{(2 \pi)^4} e^{-i p \cdot (x - y)} \mathsf{R} \frac{i R_{\widehat{N}_{\widehat{\alpha}}} p_{\widehat{N}_{\widehat{\alpha}}}}{p^2 - p_{\widehat{N}_{\widehat{\alpha}}}^2} + \cdots
	= \langle \Omega | \mathsf{R} \widehat{N}_{\widehat{\alpha}}^f (x) \overline{\widehat{N}_{\widehat{\alpha}}^i} (y) \mathsf{R} | \Omega \rangle, \ (x^0 > y^0),
\end{align}
where $\widehat{N}_{\widehat{\alpha}}^i \coloneqq (\widehat{N}_{\widehat{\alpha}}^f)^c$ and
\begin{align}
	\widehat{N}_{R \widehat{\alpha}}^f \coloneqq \mathsf{R} \widehat{N}_{\widehat{\alpha}}^f
	= \mathsf{R} \sum_\beta [(C_R^{\widehat{N}_{\widehat{\alpha}}})^{-1}]_{\widehat{\alpha} \beta} N_\beta
	= \sum_\beta [(C_R^{\widehat{N}_{\widehat{\alpha}}})^{-1}]_{\widehat{\alpha} \beta} N_{R \beta}.
\end{align}
In a similar way, we can also find from $\Delta_{LL}$
\begin{align}
	\widehat{N}_{L \widehat{\alpha}}^f \coloneqq \mathsf{L} \widehat{N}_{\widehat{\alpha}}^f
	= \mathsf{L} \sum_\beta [(C_L^{\widehat{N}_{\widehat{\alpha}}})^{-1}]_{\widehat{\alpha} \beta} N_\beta
	= \sum_\beta [(C_L^{\widehat{N}_{\widehat{\alpha}}})^{-1}]_{\widehat{\alpha} \beta} N_{R \beta}^c.
\end{align}
Furthermore, from $\Delta_{LR}$ which can written as
\begin{align}
	i (\Delta_{LR})_{\beta \alpha} (p^2) &= \sum_{\widehat{\gamma}} (C_L^{\widehat{N}_{\widehat{\gamma}}})_{\beta {\widehat{\gamma}}} \frac{i R_{\widehat{N}_{\widehat{\gamma}}}}{p^2 - p_{\widehat{N}_{\widehat{\gamma}}}^2} (C_R^{\widehat{N}_{\widehat{\gamma}}})_{\alpha {\widehat{\gamma}}} + \cdots,
\end{align}
and also from the corresponding time-ordered two-point function
\begin{align}
	\int \frac{d^4 p}{(2 \pi)^4} e^{-i p \cdot (x - y)} \mathsf{L} i \slashed{p} (\Delta_{LR})_{\beta \alpha} (p^2) \mathsf{R}
	= \langle \Omega | \mathsf{L} N_\beta (x) \overline{N_\alpha} (y) \mathsf{R} | \Omega \rangle, \ (x^0 > y^0),
\end{align}
we obtain
\begin{align}
	\int \frac{d^4 p}{(2 \pi)^4} e^{-i p \cdot (x - y)} \mathsf{L} \slashed{p} \frac{i R_{\widehat{N}_{\widehat{\alpha}}}}{p^2 - p_{\widehat{N}_{\widehat{\alpha}}}^2} + \cdots
	= \langle \Omega | \mathsf{L} \widehat{N}_{\widehat{\alpha}}^f (x) \overline{\widehat{N}_{\widehat{\alpha}}^i} (y) \mathsf{R} | \Omega \rangle, \ (x^0 > y^0).
\end{align}
Similarly, we can also find from $\Delta_{RL}$ and $\Delta_{RL}$
\begin{align}
	\int \frac{d^4 p}{(2 \pi)^4} e^{-i p \cdot (x - y)} \mathsf{R} \slashed{p} \frac{i R_{\widehat{N}_{\widehat{\alpha}}}}{p^2 - p_{\widehat{N}_{\widehat{\alpha}}}^2} + \cdots
	&= \langle \Omega | \mathsf{R} \widehat{N}_{\widehat{\alpha}}^f (x) \overline{\widehat{N}_{\widehat{\alpha}}^i} (y) \mathsf{L} | \Omega \rangle, \ (x^0 > y^0), \\
	\int \frac{d^4 p}{(2 \pi)^4} e^{-i p \cdot (x - y)} \mathsf{L} \slashed{p} \frac{i R_{\widehat{N}_{\widehat{\alpha}}}}{p^2 - p_{\widehat{N}_{\widehat{\alpha}}}^2} + \cdots
	&= \langle \Omega | \mathsf{L} \widehat{N}_{\widehat{\alpha}}^f (x) \overline{\widehat{N}_{\widehat{\alpha}}^i} (y) \mathsf{R} | \Omega \rangle, \ (x^0 > y^0).
\end{align}
Hence, the diagonalized propagator can be written as
\begin{align}
	\boxed{\int \frac{d^4 p}{(2 \pi)^4} e^{-i p \cdot (x - y)} \frac{i R_{\widehat{N}_{\widehat{\alpha}}}}{\slashed{p} - p_{\widehat{N}_{\widehat{\alpha}}}} + \cdots
	= \langle \Omega | \widehat{N}_{\widehat{\alpha}}^f (x) \overline{\widehat{N}_{\widehat{\alpha}}^i} (y) | \Omega \rangle, \ (x^0 > y^0),}
	\label{eq:CorrD}
\end{align}
where
\begin{align}
	\boxed{\widehat{N}_{\widehat{\alpha}}^f = (\widehat{N}_{\widehat{\alpha}}^i)^c
	= \sum_\beta \big[ (C_R^{\widehat{N}_{\widehat{\alpha}}})^{\minus 1} \mathsf{R} + (C_L^{\widehat{N}_{\widehat{\alpha}}})^{\minus 1} \mathsf{L} \big]_{\widehat{\alpha} \beta} N_\beta.}
\end{align}
Calculating the Fourier transform in equation \ref{eq:CorrD} in the rest frame, we can write
\begin{align}
	e^{-i p_{\widehat{N}_{\widehat{\alpha}}} (x^0 - y^0)} = e^{-i m_{\widehat{N}_{\widehat{\alpha}}} (x^0 - y^0)} e^{-(\Gamma_{\widehat{N}_{\widehat{\alpha}}} / 2) (x^0 - y^0)}
	\propto \langle \Omega | \widehat{N}_{\widehat{\alpha}}^f (x) \overline{\widehat{N}_{\widehat{\alpha}}^i} (y) | \Omega \rangle, \ (x^0 > y^0),
	\label{eq:CorrDExp}
\end{align}
\textit{i.e.}, this is the correlation function associated with the degree of freedom that propagates like a free particle until it decays.

\begin{figure}[h]
	\centering
	\includegraphics[width = 100 mm]{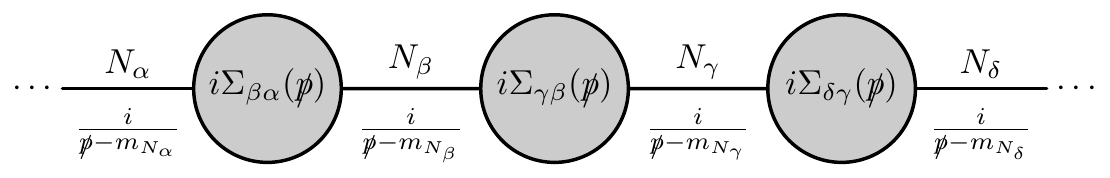}
	\caption{A non-perturbative effect is generated when $\widehat{N}_{\widehat{\alpha}}$ is on-shell.}
	\label{fig:NonPert}
\end{figure}

Now we prove $\widehat{N}_{\widehat{\alpha}}^f \neq \widehat{N}_{\widehat{\alpha}}^i$. Since equation \ref{eq:CLTCR}, \textit{i.e.}, $(C_L^{\widehat{N}_{\widehat{\alpha}}})^\mathsf{T} C_R^{\widehat{N}_{\widehat{\alpha}}} = 1 + \mathcal{O} (f^2 / 4 \pi)$, is satisfied, we can write $(C^{\widehat{N}_{\widehat{\alpha}} *}_L)^{-1} = C^{\widehat{N}_{\widehat{\alpha}} \dag}_R + \mathcal{O} (f^2 / 4 \pi)$. It follows that $(C^{\widehat{N}_{\widehat{\alpha}} *}_L)^{-1} \neq (C_R^{\widehat{N}_{\widehat{\alpha}}})^{-1}$ since $C_R^{\widehat{N}_{\widehat{\alpha}}}$ is non-unitary. Hence,
\begin{align}
	\widehat{N}_{\widehat{\alpha}}^i &= \gamma^0 \mathsf{C} (\widehat{N}_{\widehat{\alpha}}^f)^* 
	= \sum_\beta \big[ (C^{\widehat{N}_{\widehat{\alpha}} *}_L)^{\minus 1} \mathsf{R} + (C^{\widehat{N}_{\widehat{\alpha}} *}_R)^{\minus 1} \mathsf{L} \big]_{\widehat{\alpha} \beta} N_\beta
	\ \neq \ \widehat{N}_{\widehat{\alpha}}^f.
	\quad q.e.d.
\end{align}
For $\Delta m_N \lesssim \Gamma_N$, the difference between $\widehat{N}_{\widehat{\alpha}}^f$ and $\widehat{N}_{\widehat{\alpha}}^i$ can go beyond $\mathcal{O} (f^2 / 4 \pi)$. In section \ref{sec:Example}, an example in which the difference is as large as $\mathcal{O} (1)$ will be presented. This occurs due to the generic non-perturbative effect for an on-shell unstable particle. Let us consider the loop effect in figure \ref{fig:NonPert}.
For on-shell $\widehat{N}_{\widehat{\alpha}}$ where $\slashed{p} \sim m_{N_\alpha}$, the factor $\Sigma_{\beta \alpha} (\slashed{p}) / (\slashed{p} - m_{N_\alpha})$ gets highly enhanced so that the collective loop effect goes well beyond the typical perturbative correction $\mathcal{O} (f^2 / 4 \pi)$ of the theory. An on-shell stable particle does not necessarily have such a non-perturbative effect especially in the on-shell renormalization scheme where $\Sigma_{\beta \alpha} (m_{N_{\widehat{\alpha}}}) = 0$ and $\lim_{\slashed{p} \to m_{\widehat{N}_{\widehat{\alpha}}}} | \Sigma_{\beta \alpha} (\slashed{p}) / (\slashed{p} - m_{\widehat{N}_{\widehat{\alpha}}}) | \sim \mathcal{O} (f^2 / 4 \pi)$. In contrast, an unstable particle has an absorptive part in the self-energy, and it is not affected by the choice of renormalization conditions, \textit{i.e.}, $\Sigma_{\beta \alpha} (p_{\widehat{\Phi}_{\widehat{\alpha}}}^2) \sim \Sigma_{\beta \alpha} (m_{\widehat{\Phi}_{\widehat{\alpha}}}^2) \sim \mathcal{O} (f^2 / 4 \pi)$ in any renormalization scheme. As a result, the deviation of the quasiparticle from a unitary combination of $N_\alpha$ can be much larger than the typical perturbative correction. \\

In this section, we have obtained two interesting results about the particle of $\widehat{N}_{\widehat{\alpha}}$ which is the degree of freedom that propagates like a free particle until it decays in the presence of flavor mixing:
\begin{enumerate}
	\item \textit{The particle of $\widehat{N}_{\widehat{\alpha}}$ emerges as an excitation of $\widehat{N}_{\widehat{\alpha}}^i$ and ends as an excitation of $\widehat{N}_{\widehat{\alpha}}^f$ where $\widehat{N}_{\widehat{\alpha}}^f = (\widehat{N}_{\widehat{\alpha}}^i)^c \neq \widehat{N}_{\widehat{\alpha}}^i$.}
	\item \textit{The particle of $\widehat{N}_{\widehat{\alpha}}$ loses Majorana nature.}
\end{enumerate}
The first result implies that the degree of freedom should be interpreted as a \textit{quasiparticle}, \textit{i.e.}, an emergent particle dynamically generated by interactions. Since it cannot be related to a single linear combination of basis states, the Majorana condition cannot be applied. This phenomenon is similar to the generation of quasiparticles by particle-antiparticle mixing of neutral scalar particles discussed in reference \cite{QFTMixing}. The quasiparticle in that case is no longer a CP eigenstate, which is also similar to the second result mentioned above.

\section{Examples}		\label{sec:Example}
In this section, the mixing matrices and effective Yukawa couplings for two flavors will be explicitly calculated. We will discuss three different cases, two of which are in the mutually opposite limits in terms of the mass difference, \textit{i.e.}, $\Delta m_N \gg \Gamma_N$ and $\Delta m_N \ll \Gamma_N$. Those cases can be analytically studied, which will help understand the intermediate case $\Delta m_N \sim \Gamma_N$. The case of $\Delta m_N \sim \Gamma_N$ can only be numerically analyzed, and an example with large non-unitary mixing will be presented. Beyond two flavors, it is in general complicated to carry out calculations in an analytic way, but we can follow the same steps as in the two-flavor case discussed here. \\

For two flavors of RH Majorana neutrinos, the matrices $A$ and $B$ defined by equations \ref{eq:A} and \ref{eq:B} are explicitly written as
\begin{align}
	A (p^2) &= \left( \begin{array}{cc} 1 + (\Sigma_R)_{11} & (\Sigma_R)_{12} \\
			(\Sigma_R)_{21} & 1 + (\Sigma_R)_{22} \end{array} \right), \\
	B (p^2) &= M_N^\frac{1}{2} A^{-1} M_N^\frac{1}{2}
		= \frac{1}{\text{det} [A]} \left( \begin{array}{cc} m_{N_1} [(1 + (\Sigma_R)_{22}] & \minus \sqrt{m_{N_1} m_{N_2}} (\Sigma_R)_{12} \\
				\minus \sqrt{m_{N_1} m_{N_2}} (\Sigma_R)_{21} & m_{N_2} [1 + (\Sigma_R)_{11}] \end{array} \right),
\end{align}
where
\begin{align}
	\text{det} [A (p^2)] = [1 + (\Sigma_R)_{11}] [1 + (\Sigma_R)_{22}] - (\Sigma_R)_{12} (\Sigma_R)_{21}.
\end{align}
Hence,
\begin{align}
	B^\mathsf{T} (p^2) B (p^2)
		= \frac{1}{\{\text{det} [A]\}^2} \left( \begin{array}{cc} a_1 & b_{12} \\ b_{12} & a_2 \end{array} \right),
\end{align}
where
\begin{align}
	a_1 (p^2) &\coloneqq m_{N_1}^2 [1 + (\Sigma_R)_{22}]^2 + m_{N_1} m_{N_2} [(\Sigma_R)_{21}]^2, \\
	a_2 (p^2) &\coloneqq m_{N_2}^2 [1 + (\Sigma_R)_{11}]^2 + m_{N_1} m_{N_2} [(\Sigma_R)_{12}]^2, \\
	b_{12} (p^2) &\coloneqq \minus \sqrt{m_{N_1} m_{N_2}} \big\{ m_{N_1} [1 + (\Sigma_R)_{22}] (\Sigma_R)_{12} + m_{N_2} [1 + (\Sigma_R)_{11}] (\Sigma_R)_{21} \big\}.
\end{align}
Using the identity
\begin{align}
	\{\text{det} [A (p^2)]\}^2 = \frac{a_1 a_2 - b_{12}^2}{m_{N_1}^2 m_{N_2}^2},
\end{align}
we can write
\begin{align}
	B^\mathsf{T} (p^2) B (p^2)
	= \frac{m_{N_1}^2 m_{N_2}^2}{a_1 a_2 - b_{12}^2}
		\left( \begin{array}{cc} a_1 & b_{12} \\ b_{12} & a_2 \end{array} \right).
\end{align}
The eigenvalues of $B^\mathsf{T} B$ as functions of $p^2$ are given by
\begin{align}
	P_1^2 (p^2) = \frac{m_{N_1}^2 m_{N_2}^2}{a_1 a_2 - b_{12}^2} \lambda_1, \qquad
	P_2^2 (p^2) = \frac{m_{N_1}^2 m_{N_2}^2}{a_1 a_2 - b_{12}^2} \lambda_2
\end{align}
where
\begin{align}
	\lambda_1 (p^2) &\coloneqq \frac{1}{2} \bigg[ (a_1 + a_2) - \sqrt{(a_2 - a_1)^2 + 4 b_{12}^2} \bigg], \\
	\lambda_2 (p^2) &\coloneqq \frac{1}{2} \bigg[ (a_1 + a_2) + \sqrt{(a_2 - a_1)^2 + 4 b_{12}^2} \bigg].
\end{align}
Here, the square-root of a complex number is defined by its principal branch: $\arg[\sqrt{z}] \in (-\pi / 2, \pi /2]$ for a complex number $z$. The orthogonal matrix $O_L$ which diagonalizes $B^\mathsf{T} B$ by $P^2 (p^2) = O_L^\mathsf{T} B^\mathsf{T} B O_L$ is given by
\begin{align}
	\boxed{O_L (p^2) = \frac{\sqrt{b_{12}^2} / b_{12}}{\sqrt{(a_1 - \lambda_1)^2 + b_{12}^2}}
		\left( \begin{array}{cc} b_{12} & a_1 - \lambda_1 \\ \minus (a_1 - \lambda_1) & b_{12} \end{array} \right).}
\end{align}
Here, the phase factor $\sqrt{b_{12}^2} / b_{12}$ is introduced to set $(O_L)_{\alpha \alpha} = 1$ when $a_1 - \lambda_1 = 0$. We can also obtain the other mixing matrices from $O_R (p^2) = B O_L P^{-1}$, $C_R (p^2) = M_N^{-\frac{1}{2}} O_R M_N^\frac{1}{2}$, and $C_L (p^2) = M_N^{-\frac{1}{2}} O_L M_N^\frac{1}{2}$.

\subsection{Large mass difference}
Let us first consider a large mass difference, which has been denoted by $\Delta m_N \gg \Gamma_N$ and can be more accurately written as
\begin{align}
	\boxed{|m_{N_\beta} - m_{N_\alpha}| \gg m_{N_\beta} \mathcal{O} (f^2 / 4 \pi).}
\end{align}
In addition, for convenience, we assume that the mass difference is not too large:
\begin{align}
	\boxed{m_{N_\alpha} \gg m_{N_\beta} \mathcal{O} (f^2 / 4 \pi).}
	\label{eq:ConLarge}
\end{align}
In this section, we will determine the order of perturbation in terms of $\Sigma_R (p^2)$ rather than $f^2 / 4 \pi$. Up to $\mathcal{O} (\Sigma^2)$,
\begin{align}
	\bigg| \frac{4 b_{12}^2}{(a_2 - a_1)^2} \bigg|
	= \frac{4 m_{N_1} m_{N_2} [m_{N_1}^2 \mathcal{O} (\Sigma^2) + m_{N_2}^2 \mathcal{O} (\Sigma^2)]}{(m_{N_2}^2 - m_{N_1}^2)^2}
	\ll 1,
\end{align}
and we can therefore use $4 b_{12}^2 / (a_2 - a_1)^2$ as a small expansion parameter.

The eigenvalues of $\{\text{det} [A]\}^2 B^\mathsf{T} B$ up to $\mathcal{O} (\Sigma^2)$ are given by
\begin{align}
	\lambda_1 (p^2) &= \frac{1}{2} \bigg[ (a_1 + a_2) - (a_2 - a_1) - \frac{2 b_{12}^2}{a_2 - a_1} \bigg]
	= a_1 - \frac{b_{12}^2}{m_{N_2}^2 - m_{N_1}^2}, \\
	\lambda_2 (p^2) &= \frac{1}{2} \bigg[ (a_1 + a_2) + (a_2 - a_1) + \frac{2 b_{12}^2}{a_2 - a_1} \bigg]
	= a_2 + \frac{b_{12}^2}{m_{N_2}^2 - m_{N_1}^2},
\end{align}
where we have assumed $\sqrt{(a_2 - a_1)^2} = a_2 - a_1$. The other possibility is $\sqrt{(a_2 - a_1)^2} = a_1 - a_2$, in which case we may rename $\lambda_1$ to $\lambda_2$ and vice versa. Then, up to $\mathcal{O} (\Sigma)$
\begin{align}
	P_1^2 (p^2) &= \frac{m_{N_1}^2 m_{N_2}^2}{a_1 a_2 - b_{12}^2} \lambda_1
		= \frac{m_{N_1}^2 m_{N_2}^2}{a_2}
		= m_{N_1}^2 [1 - 2 (\Sigma_R)_{11}], \\
	P_2^2 (p^2) &= \frac{m_{N_1}^2 m_{N_2}^2}{a_1 a_2 - b_{12}^2} \lambda_2
		= \frac{m_{N_1}^2 m_{N_2}^2}{a_1}
		= m_{N_1}^2 [1 - 2 (\Sigma_R)_{22}],
\end{align}
where we have used condition \ref{eq:ConLarge}. Hence, we can write up to $\mathcal{O} (\Sigma)$
\begin{align}
	\boxed{P_{\widehat{\alpha}} (p^2) = m_{N_\alpha} [1 - (\Sigma_R)_{\alpha \alpha} (p^2)],}
\end{align}
\textit{i.e.},
\begin{align}
	(\widehat{\Sigma}_R)_{\widehat{\alpha}} (p^2) = (\Sigma_R)_{\alpha \alpha} (p^2).
\end{align}
The complex mass is given by
\begin{align}
	p_{\widehat{N}_{\widehat{\alpha}}} = P_{\widehat{\alpha}} (p_{\widehat{N}_{\widehat{\alpha}}}^2)
	= m_{N_\alpha} [1 - (\Sigma_R)_{\alpha \alpha} (p_{\widehat{N}_{\widehat{\alpha}}}^2)]
	= m_{\widehat{N}_{\widehat{\alpha}}} - i \frac{\Gamma_{\widehat{N}_{\widehat{\alpha}}}}{2},
\end{align}
and thus
\begin{align}
	\boxed{\text{Re} [(\Sigma_R)_{\alpha \alpha} (p_{\widehat{N}_{\widehat{\alpha}}}^2)] = \frac{m_{\widehat{N}_{\widehat{\alpha}}}}{m_{N_\alpha}} - 1, \qquad
	\text{Im} [(\Sigma_R)_{\alpha \alpha} (p_{\widehat{N}_{\widehat{\alpha}}}^2)] = \frac{\Gamma_{\widehat{N}_{\widehat{\alpha}}}}{2 m_{N_\alpha}}.}
\end{align}

Now we calculate $C_A (p^2)$ and effective Yukawa couplings. Up to $\mathcal{O} (\Sigma^2)$, we can write
\begin{align}
	a_1 - \lambda_1 = \frac{b_{12}^2}{m_{N_2}^2 - m_{N_1}^2}, \qquad
	(a_1 - \lambda_1)^2 + b_{12}^2 = b_{12}^2,
\end{align}
and thus up to $\mathcal{O} (\Sigma)$
\begin{align}
	O_L (p^2) &= \left( \begin{array}{cc} 1 & \frac{a_1 - \lambda_1}{b_{12}} \\
			\minus \frac{a_1 - \lambda_1}{b_{12}} & 1 \end{array} \right)
		= \left( \begin{array}{cc} 1 & \frac{b_{12}}{m_{N_2}^2 - m_{N_1}^2} \\
			\minus \frac{b_{12}}{m_{N_2}^2 - m_{N_1}^2} & 1 \end{array} \right) \nonumber \\
		&= \left( \begin{array}{cc} 1 & \minus \frac{\sqrt{m_{N_1} m_{N_2}} [m_{N_1} (\Sigma_R)_{12} + m_{N_2} (\Sigma_R)_{21}]}{m_{N_2}^2 - m_{N_1}^2} \\
			\frac{\sqrt{m_{N_1} m_{N_2}} [m_{N_1} (\Sigma_R)_{12} + m_{N_2} (\Sigma_R)_{21}]}{m_{N_2}^2 - m_{N_1}^2} & 1 \end{array} \right), \\
	O_R (p^2) &= B O_L P^{-1} \nonumber \\
		&= \left( \begin{array}{cc} 1 & \minus \frac{\sqrt{m_{N_1} m_{N_2}} [m_{N_2} (\Sigma_R)_{12} + m_{N_1} (\Sigma_R)_{21}]}{m_{N_2}^2 - m_{N_1}^2} \\
			\frac{\sqrt{m_{N_1} m_{N_2}} [m_{N_2} (\Sigma_R)_{12} + m_{N_1} (\Sigma_R)_{21}]}{m_{N_2}^2 - m_{N_1}^2} & 1 \end{array} \right).
\end{align}
We therefore obtain
\begin{empheq}[box=\fbox]{align}
	C_R (p^2) &= M_N^{-\frac{1}{2}} O_R M_N^\frac{1}{2}
		= \left( \begin{array}{cc} 1 & \minus \frac{m_{N_2} [m_{N_2} (\Sigma_R)_{12} + m_{N_1} (\Sigma_R)_{21}]}{m_{N_2}^2 - m_{N_1}^2} \\
			\frac{m_{N_1} [m_{N_2} (\Sigma_R)_{12} + m_{N_1} (\Sigma_R)_{21}]}{m_{N_2}^2 - m_{N_1}^2} & 1 \end{array} \right), \\
	C_L (p^2) &= M_N^{-\frac{1}{2}} O_L M_N^\frac{1}{2}
		= \left( \begin{array}{cc} 1 & \minus \frac{m_{N_2} [m_{N_1} (\Sigma_R)_{12} + m_{N_2} (\Sigma_R)_{21}]}{m_{N_2}^2 - m_{N_1}^2} \\
			\frac{m_{N_1} [m_{N_1} (\Sigma_R)_{12} + m_{N_2} (\Sigma_R)_{21}]}{m_{N_2}^2 - m_{N_1}^2} & 1 \end{array} \right).
\end{empheq}
The effective Yukawa couplings incorporating the vertex-loop correction as well are defined by equation \ref{eq:Feff}, and they are given by
\begin{empheq}[box=\fbox]{align}
	\widehat{F}_{i \widehat{\alpha}} &= (D_R^{\widehat{N}_{\widehat{\alpha}}} f C_R^{\widehat{N}_{\widehat{\alpha}}})_{i \alpha} \nonumber \\
		&= f_{i \alpha} + m_{N_\alpha} V_{RL} (m_{N_\alpha}^2) + f_{i \beta} \frac{m_{N_\alpha} [m_{N_\beta} (\Sigma_R)_{\alpha \beta} (m_{N_\alpha}^2) + m_{N_\alpha} (\Sigma_R)_{\beta \alpha} (m_{N_\alpha}^2)]}{m_{N_\beta}^2 - m_{N_\alpha}^2},
		\label{eq:FeffLarge} \\
	\widehat{F}^c_{i \widehat{\alpha}} &= (D_L^{\widehat{N}_{\widehat{\alpha}}} f^* C_L^{\widehat{N}_{\widehat{\alpha}}})_{i \alpha} \nonumber \\
		&= f_{i \alpha}^* + m_{N_\alpha} V_{LR} (m_{N_\alpha}^2) + f_{i \beta}^* \frac{m_{N_\alpha} [m_{N_\beta} (\Sigma_R)_{\beta \alpha} (m_{N_\alpha}^2) + m_{N_\alpha} (\Sigma_R)_{\alpha \beta} (m_{N_\alpha}^2)]}{m_{N_\beta}^2 - m_{N_\alpha}^2},
		\label{eq:FceffLarge}
\end{empheq}
where $\beta \neq \alpha$ and we have used $\Sigma_R (p_{\widehat{N}_{\widehat{\alpha}}}^2) = \Sigma_R (m_{\widehat{N}_{\widehat{\alpha}}}^2) = \Sigma_R (m_{N_\alpha}^2)$ which is correct up to $\mathcal{O} (\Sigma)$. Note that equations \ref{eq:FeffLarge} and \ref{eq:FceffLarge} are the effective Yukawa couplings that are consistent with the oldest expression of the CP asymmetry \cite{CPVLepto} in the literature, if the vertex loops are neglected and the decay widths of $\widehat{N}_{\widehat{\alpha}}$ are calculated from
\begin{align}
	\Gamma (\widehat{N}_{\widehat{\alpha}} \to L_i \phi) = \frac{m_{\widehat{N}_{\widehat{\alpha}}}}{16 \pi} \widehat{f}_{i \widehat{\alpha}}^* \widehat{f}_{i \widehat{\alpha}}, \qquad
	\Gamma (\widehat{N}_{\widehat{\alpha}} \to L_i^c \phi^*) = \frac{m_{\widehat{N}_{\widehat{\alpha}}}}{16 \pi} \widehat{f}_{i \widehat{\alpha}}^{c *} \widehat{f}_{i \widehat{\alpha}}^c.
\end{align}
These expression of decay widths are, however, invalid since they are obtained by regarding the quasiparticle as an external state. The quasiparticle is dynamically generated by interactions, and its property can be correctly studied only when it is treated as an intermediate state. In section \ref{sec:Example3}, we will see that the total decay width calculated from $\sum_i \big[ \Gamma (\widehat{N}_{\widehat{\alpha}} \to L_i \phi) + \Gamma (\widehat{N}_{\widehat{\alpha}} \to L_i^c \phi^*) \big]$ does not give the correct value.

All the other expressions of effective Yukawa couplings presented in the literature are sorts of some perturbative corrections to equations \ref{eq:FeffLarge} and \ref{eq:FceffLarge}. In particular, the focus has been to find the next-order correction, \textit{i.e.}, the regulator, to the denominators of their $\mathcal{O} (f^3 / 4 \pi)$ terms. However, the correction for a small mass difference is actually as large as the leading order terms due to the non-perturbative effects mentioned in section \ref{sec:Quasi}. We will explicitly see such cases in the following examples.

\subsection{Extremely small mass difference}
When the mass difference is in the opposite limit:
\begin{align}
	|m_{N_\beta} - m_{N_\alpha}| \ll m_{N_\alpha} \mathcal{O} (f^2 / 4 \pi),
\end{align}
which has been denoted by $\Delta m_N \ll \Gamma_N$, there does not exist a useful expansion parameter in general. For example,
\begin{align}
	\bigg| \frac{(a_1 - a_2)^2}{4 b_{12}^2} \bigg|
	= \bigg| \frac{[m_{N_2}^2 \mathcal{O} (\Sigma) + m_{N_1}^2 \mathcal{O} (\Sigma)]^2}{m_{N_1} m_{N_2} [m_{N_1} \mathcal{O} (\Sigma) + m_{N_2} \mathcal{O} (\Sigma)]^2} \bigg|
\end{align}
may or may not be smaller than one. Hence, to investigate this case in an analytic way, we consider an extreme case of
\begin{align}
	\boxed{N_1 \neq N_2, \qquad
	m_N \coloneqq m_{N_1} = m_{N_2}, \qquad
	f_i \coloneqq f_{i1} = f_{i2}.}
\end{align}
For two flavors to be theoretically distinguishable, \textit{i.e.}, $N_1 \neq N_2$, the masses or Yukawa couplings will have to be at least slightly different, and the conditions should be considered to be correct only up to the working precision. In this case, we have
\begin{align}
	(\Sigma_{0 R})_{\beta \alpha} (p^2) = \sum_i \frac{|f_i|^2}{16 \pi^2} \bigg[ \minus \log{\bigg( \frac{|p^2|}{\Lambda^2} \bigg)} + i (\pi  - \arg{[p^2]}) \bigg].
\end{align}
Choosing every component of $\delta_N$ to be identical, we can make the renormalized self-energy satisfy
\begin{align}
	\Sigma_R' (p^2) \coloneqq (\Sigma_R)_{\beta \alpha} (p^2).
\end{align}
Defining
\begin{align}
	a (p^2) &\coloneqq a_1 = a_2 = m_N^2 [(1 + \Sigma_R')^2 + (\Sigma_R')^2], \\
	b (p^2) &\coloneqq b_{12} = \minus 2 m_N^2 (1 + \Sigma_R') \Sigma_R',
\end{align}
we can write
\begin{align}
	\lambda_1 (p^2) = a - b = m_N^2 (1 + 2 \Sigma_R')^2, \qquad
	\lambda_2 (p^2) = a + b = m_N^2,
\end{align}
where $\sqrt{b^2} = b$ has been assumed. If $\sqrt{b^2} = \minus b$, we may rename $\lambda_1$ to $\lambda_2$ and vice versa. Then,
\begin{align}
	P_1^2 (p^2) &= \frac{m_{N_1}^2 m_{N_2}^2}{a_1 a_2 - b_{12}^2} \lambda_1
		= \frac{m_N^4}{a + b}
		= m_N^2, \\
	P_2^2 (p^2) &= \frac{m_{N_1}^2 m_{N_2}^2}{a_1 a_2 - b_{12}^2} \lambda_2
		= \frac{m_N^4}{a - b}
		= m_N^2 (1 - 2 \Sigma_R')^2.
\end{align}
Up to $\mathcal{O} (\Sigma)$, we can write
\begin{align}
	\boxed{P_1 (p^2) = m_N, \qquad
	P_2 (p^2) = m_N [1 - 2 \Sigma_R' (p^2)].}
\end{align}
\textit{i.e.},
\begin{align}
	(\widehat{\Sigma}_R)_1 (p^2) = 0, \qquad
	(\widehat{\Sigma}_R)_2 (p^2) = 2 \Sigma_R' (p^2).
\end{align}
The complex poles are now written as
\begin{align}
	p_{\widehat{N}_1} &= P_1 (p_{\widehat{N}_1}^2)
		= m_N = m_{\widehat{N}_1} - i \frac{\Gamma_{\widehat{N}_1}}{2}, \\
	p_{\widehat{N}_2} &= P_2 (p_{\widehat{N}_2}^2)
		= m_N [1 - 2 \Sigma_R' (p_{\widehat{N}_2}^2)]
		= m_{\widehat{N}_2} - i \frac{\Gamma_{\widehat{N}_2}}{2},
\end{align}
and thus
\begin{align}
	\boxed{m_{\widehat{N}_1} = m_N, \quad
	\Gamma_{\widehat{N}_1} = 0, \quad
	\text{Re} [\Sigma_R' (p_{\widehat{N}_2}^2)] = \frac{1}{2} \bigg( \frac{m_{\widehat{N}_2}}{m_N} - 1 \bigg), \quad
	\text{Im} [\Sigma_R' (p_{\widehat{N}_2}^2)] = \frac{\Gamma_{\widehat{N}_2}}{4 m_N}.}
\end{align}
Note that $\widehat{N}_1$ is a field of a stable particle, which is dynamically generated from the apparently unstable particles by exact cancellation of self-energies.

Moreover, using
\begin{align}
	a_1 - \lambda_1 = b,
\end{align}
we can write
\begin{align}
	O_L (p^2) = \frac{1}{\sqrt{2}} \left( \begin{array}{cc} 1 & 1 \\ \minus 1 & 1 \end{array} \right), \qquad
	O_R (p^2) = B O_L P^{-1}
		= \frac{1}{\sqrt{2}} \left( \begin{array}{cc} 1 & 1 \\ \minus 1 & 1 \end{array} \right).
\end{align}
Hence,
\begin{align}
	\boxed{C_R (p^2) = M_N^{-\frac{1}{2}} O_R M_N^\frac{1}{2}
		= \frac{1}{\sqrt{2}} \left( \begin{array}{cc} 1 & 1 \\
			\minus 1 & 1 \end{array} \right), \qquad
	C_L (p^2) = M_N^{-\frac{1}{2}} O_L M_N^\frac{1}{2}
		= \frac{1}{\sqrt{2}} \left( \begin{array}{cc} 1 & 1 \\
			\minus 1 & 1 \end{array} \right).}
\end{align}
The effective Yukawa couplings are given by
\begin{empheq}[box=\fbox]{align}
	&\widehat{f}_{i1} = (f C^{\widehat{N}_1}_R)_{i1}
		= \frac{1}{\sqrt{2}} (1 - 1) f_i = 0,
	&&\widehat{f}_{i2} = (f C^{\widehat{N}_2}_R)_{i2}
		= \frac{1}{\sqrt{2}} (1 + 1) f_i = \sqrt{2} f_i, \\
	&\widehat{f}^c_{i1} = (f^* C^{\widehat{N}_1}_L)_{i1}
		= \frac{1}{\sqrt{2}} (1 - 1) f_i^* = 0,
	&&\widehat{f}^c_{i2} = (f^* C^{\widehat{N}_2}_L)_{i2}
		= \frac{1}{\sqrt{2}} (1 + 1) f_i^* = \sqrt{2} f_i^*.
\end{empheq}

Note that $O_A (p^2)$ and $C_A (p^2)$ are maximal-mixing unitary matrices. In fact, before any calculation and derivation as done here, we could have gone to a new basis using those unitary matrices. The mass matrix is still diagonal in the new basis, and the effective Yukawa coupligs calculated here are just normal Yukawa couplings there. In that case, not only two RH neutrinos are completely decoupled because the Yukawa coupling matrix is diagonal, but also one of them is completely stable because it is a free field without any interaction. In other words, the basis we chose here is actually an inconvenient one which requires a more complicated analysis. The analysis as is done here, however, is still illuminating since this is an extreme case which allows an analytic approach and it shows that $C_A (p^2)$ is a large mixing matrix when the mass difference is small. When the mass difference is small but not so extreme as in this case, we cannot simply change the basis because the mass matrix in the new basis would no longer be diagonal. The current case with identical masses is unique because the mass matrix is still diagonal after change of basis by $C_A (p^2)$. In general, the requirement that the mass matrix be diagonal eliminates the freedom to change a basis, and the analysis needs to be done as discussed here.

\subsection{Small mass difference}		\label{sec:Example3}
When the mass difference is not so extremely small and satisfies
\begin{align}
	|m_{N_\beta} - m_{N_\alpha}| \sim m_{N_\alpha} \mathcal{O} (f^2 / 4 \pi),
\end{align}
\textit{i.e.}, $\Delta m_N \sim \Gamma_N$, we do not generally have a useful expansion parameter similarly to the case of an extremely small mass difference, and have to depend on numerical calculation to find the mixing matrices and effective Yukawa couplings. Here, we discuss an example where $C_A (p^2)$ are non-unitary large-mixing matrices.

We choose the masses of RH neutrinos
\begin{align}
	m_{N_1} = 1~\text{TeV}, \qquad
	m_{N_2} - m_{N_1} = 10^{-9.2}~\text{TeV}
\end{align}
and a Yukawa coupling matrix
\begin{align}
	f = \left( \begin{array}{cc} 10^{-0.1} e^{i 0.2 \pi} & 10^{-0.2} e^{-i 0.8 \pi} \\ e^{i 0.1 \pi} & e^{i 1.2 \pi} \end{array} \right) \cdot 10^{-4}.
\end{align}
Using the expression of the self-energy given by equation \ref{eq:SEexp}:
\begin{align}
	(\Sigma_R)_{\beta \alpha} (p^2) = \sum_i \frac{f_{i \beta}^* f_{i \alpha}}{16 \pi^2} \bigg[ \minus \log{\bigg( \frac{p^2}{m_{N_\beta} m_{N_\alpha}} \bigg)} + i \pi \bigg],
\end{align}
we find
\begin{align}
	\Sigma_R (m_{N_\alpha}^2) &= \left( \begin{array}{cc} 3.24469 i & 0.61477 - 2.88915 i \\ \minus 0.61477 - 2.88915 i & 2.78145 i \end{array} \right) \cdot 10^{-10}.
\end{align}
The matrices $A (m_{N_\alpha}^2)$ and $B (m_{N_\alpha}^2)$ can be found from $\Sigma_R (m_{N_\alpha}^2)$, and $O_L (m_{N_\alpha}^2)$ is obtained by finding the orthogonal mixing matrix which diagonalizes $B^\mathsf{T} (m_{N_\alpha}^2) B (m_{N_\alpha}^2)$. The complex pole $p_{\widehat{N}_{\widehat{\alpha}}}^2 = P^2 (m_{N_\alpha}^2)$ is one of the eigenvalues of $B^\mathsf{T} (m_{N_\alpha}^2) B (m_{N_\alpha}^2)$:
\begin{align}
	p_{\widehat{N}_{\widehat{\alpha}}}^2 = [O_L^\mathsf{T} (m_{N_\alpha}^2) B^\mathsf{T} (m_{N_\alpha}^2) B (m_{N_\alpha}^2) O_L (m_{N_\alpha}^2)]_{\alpha \alpha}.
\end{align}
We can find the other mixing matrices from $O_R (m_{N_\alpha}^2) = B (m_{N_\alpha}^2) O_L (m_{N_\alpha}^2) P^{-1} (m_{N_\alpha}^2)$, $C_R^{\widehat{N}_{\widehat{\alpha}}} = M_N^{-\frac{1}{2}} O_R (m_{N_\alpha}^2) M_N^\frac{1}{2}$, and $C_L^{\widehat{N}_{\widehat{\alpha}}} = M_N^{-\frac{1}{2}} O_L (m_{N_\alpha}^2) M_N^\frac{1}{2}$. The complex masses $p_{\widehat{N}_{\widehat{\alpha}}}$ are found to be
\begin{align}
	p_{\widehat{N}_1} &= 1 + (1.79767 - 3.55150 i) \cdot 10^{-10}~\text{TeV}, \\
	p_{\widehat{N}_2} &= 1 + (\minus 1.79767 - 2.47464 i) \cdot 10^{-10}~\text{TeV},
\end{align}
\textit{i.e.},
\begin{alignat}{2}
	&m_{\widehat{N}_1} = 1 + 1.79767 \cdot 10^{-10}~\text{TeV}, \qquad
	&&\Gamma_{\widehat{N}_1} = 7.10299 \cdot 10^{-10}~\text{TeV},
		\label{eq:CMass1} \\
	&m_{\widehat{N}_2} = 1 - 1.79767 \cdot 10^{-10}~\text{TeV}, \qquad
	&&\Gamma_{\widehat{N}_2} = 4.94927 \cdot 10^{-10}~\text{TeV}.
		\label{eq:CMass2}
\end{alignat}
In addition, $C_A^{\widehat{N}_{\widehat{\alpha}}}$ are given by
\begin{align}
	C_R^{\widehat{N}_{\widehat{\alpha}}} = C_L^{\widehat{N}_{\widehat{\alpha}}} &= \left( \begin{array}{cc} 1.24519 - 0.130381 i & 0.213356 + 0.760928 i \\ \minus 0.213356 - 0.760928 i & 1.24519 - 0.130381 i \end{array} \right), \\
	C_R^{\widehat{N}_{\widehat{\alpha}}} - C_L^{\widehat{N}_{\widehat{\alpha}}} &= \left( \begin{array}{cc} 1.31165 + 4.67795 i & \minus 7.65503 + 0.801543 i \\ 7.65503 - 0.801543 i & 1.31165 + 4.67795 i \end{array} \right) \cdot 10^{-11}.
\end{align}
The deviation of $C_A^{\widehat{N}_{\widehat{\alpha}}}$ from unitarity is as large as $\mathcal{O} (1)$:
\begin{align}
	(C_A^{\widehat{N}_{\widehat{\alpha}}})^\dag C_A^{\widehat{N}_{\widehat{\alpha}}} = \left( \begin{array}{cc} 2.19202 & 1.95063 i \\ \minus 1.95063 i & 2.19202 \end{array} \right)
		+ \left( \begin{array}{cc} 0 & 1.05028 \\ 1.05028 & 0 \end{array} \right) \cdot 10^{-10},
\end{align}
In addition, they indeed satisfy $(C_L^{\widehat{N}_{\widehat{\alpha}}})^\mathsf{T} C_R^{\widehat{N}_{\widehat{\alpha}}} = 1 + \mathcal{O} (f^2 / 4 \pi)$ as implied by equation \ref{eq:CLTCR}:
\begin{align}
	(C_L^{\widehat{N}_{\widehat{\alpha}}})^\mathsf{T} C_R^{\widehat{N}_{\widehat{\alpha}}} = 1 + \left( \begin{array}{cc} \minus 2.04870 i & 1.68746 + 5.80278 i \\ 2.91700 + 5.80278 i & 2.04870 i \end{array} \right) \cdot 10^{-10}.
\end{align}
The assoicated effective Yukawa couplings are given by
\begin{align}
	\widehat{f} &= \left( \begin{array}{cc} 0.687768 + 0.965131 i & \minus 0.902131 + 0.193361 i \\ 0.949879 + 1.00180 i & -1.11624 + 0.163194 i \end{array} \right) \cdot 10^{-4}, \\
	\widehat{f}^c &= \left( \begin{array}{cc} 1.13043 - 0.355864 i & \minus 0.0948769 + 0.917729 i \\ 1.76382 - 0.0185878 i & \minus 0.492687 + 1.49514 i \end{array} \right) \cdot 10^{-4}.
\end{align}

As mentioned in section \ref{sec:Diag}, the effective Yukawa couplings can be regarded as ordinary Yukawa couplings for many practical purposes. However, it is not allowed when such a prescription implies the quasiparticle should be treated like an asymptotic state. For example, the formula of the total decay width commonly used in the literature \cite{CPMajDec, CPResLepto, ResLepto, CPVMaj, FlavCovResLepto} is given by
\begin{align}
	\Gamma_{\widehat{N}_{\widehat{\alpha}}} = \frac{m_{\widehat{N}_{\widehat{\alpha}}}}{16 \pi} \sum_i (\widehat{f}_{i \widehat{\alpha}}^* \widehat{f}_{i \widehat{\alpha}} + \widehat{f}_{i \widehat{\alpha}}^{c *} \widehat{f}_{i \widehat{\alpha}}^c),
\end{align}
and it is obtained by regarding $\widehat{N}_{\widehat{\alpha}}$ as an external field. This formula gives $\Gamma_{\widehat{N}_1} = 1.55699 \cdot 10^{-9}$ TeV and $\Gamma_{\widehat{N}_2} = 1.08489 \cdot 10^{-9}$ TeV, which are indeed different from the values in equations \ref{eq:CMass1} and \ref{eq:CMass2}.

Since quasiparticles are degrees of freedom dynamically generated by interactions, they cannot be treated like external states and their properties can only be studied from the scattering mediated by on-shell quasiparticles. Such a method to calculate the decay widths of quasiparticles and basis states in the case of particle-antiparticle mixing is developed in reference \cite{QFTMixing}. In a follow-up paper \cite{MajMixingII}, a similar method will be developed for the flavor mixing of heavy Majorana particles, and the effect of CP violation will be studied.

\section{Conclusion}
In summary, we have studied how to handle the mixing of multiple flavors of heavy Majorana neutrinos, and discussed the physics behind it. Since the on-shell renormalization scheme or complex mass scheme cannot be applied to any mass difference of heavy Majorana neutrinos, we have carefully discussed the mass and field-strength renormalization step-by-step. In order to identify the propagating particle as well as to calculate its pole mass and total decay width, we have examined the diagonalization of the resummed propagator. The diagonalization procedure presented here is exact at least up to the one-loop order in the self-energy, and thus allows an attentive study for any mass difference. We have seen that, for a small mass difference, the mixing matrices from the basis states of heavy Majorana particles to the physical degrees of freedom require large non-unitary mixing among flavors, and it is caused by the generic non-perturbative effect in an on-shell unstable particle. We have also identified the physical degree of freedom associated with each component of the diagonalized propagator, and have shown that it cannot be expressed as a single linear combination of basis states. Hence, it should be interpreted as a quasiparticle, \textit{i.e.}, an emergent particle dynamically generated by interactions. Since the Majorana condition cannot be applied to it, it must lose Majorana nature. In follow-up papers, the discussion will be continued to obtain the decay widths of quasiparticles and basis states, and the CP asymmetry in the decays of heavy Majorana neutrinos will be derived. Its application to leptogenesis will also be studied.

\section*{Acknowledgement}
This work was supported by the National Center for Theoretical Sciences, Hsinchu.

\appendix

\section{Calculation trick for diagrams with Majorana-type propagators}		\label{sec:TrickMaj}
Calculating an $S$-matrix element involving propagators of Majorana fields is complicated in general due to the presence of three different types of propagators as mentioned in section \ref{sec:BasicMaj}. Moreover, charge conjugation operators usually appear in the expression of an $S$-matrix element, and they require careful tracking of spinor indices for calculation. Here, we discuss a trick which simplifies the calculation of an $S$-matrix element with Majorana fields such that it is no more complicated than the calculation of an $S$-matrix element only with Dirac fields. The strategy is that, for a given process involving propagators of Majorana fields, we appropriately change the forms of Lagrangian terms such that only the \textit{Dirac-type} propagators appear in the corresponding $S$-matrix element and the charge conjugation operators are all absorbed into the existing fields.

\begin{figure}[h]
	\centering
	\includegraphics[width = 36 mm]{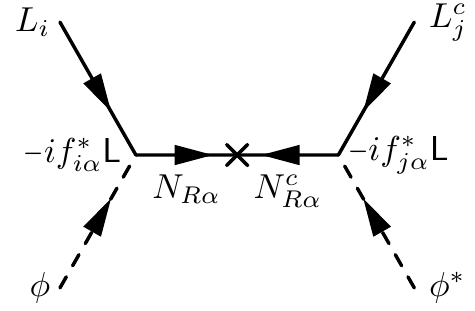}
	\caption{Tree-level contribution to $L_i \phi \to L_j^c \phi^*$.}
	\label{fig:LPLcPcTree}
\end{figure}
As a specific example, let us consider the scattering $L_i \phi \to L_j^c \phi^*$ whose tree-level diagram is given in figure \ref{fig:LPLcPcTree}. The associated Lagrangian interaction term is
\begin{align}
	\minus f_{i \alpha}^* \overline{N_\alpha} \widetilde{\phi}^\dag \mathsf{L} L_i,
\end{align}
and we claim that it can be rewritten as
\begin{align}
	\boxed{\minus f_{i \alpha}^* \overline{N_\alpha} \widetilde{\phi}^\dag \mathsf{L} L_i = f_{i \alpha}^* \overline{L_i^c} \widetilde{\phi}^* \mathsf{L} N_\alpha.}
	\label{eq:LTrick}
\end{align}
Similarly, its Hermitian conjugate term can be rewritten as
\begin{align}
	\boxed{\minus f_{i \alpha} \overline{L_i} \widetilde{\phi} \mathsf{R} N_\alpha = f_{i \alpha} \overline{N_\alpha} \widetilde{\phi}^\mathsf{T} \mathsf{R} L_i^c.}
\end{align}
In order to derive equation \ref{eq:LTrick}, first note that 
\begin{align}
	\overline{\psi} &= \psi^\dag \gamma^0
	= [\mathsf{C}^{-1} \gamma^0 (\gamma^0 \mathsf{C} \psi^*)]^\mathsf{T} \gamma^0
	= (\psi^c)^\mathsf{T} (\gamma^0)^\mathsf{T} (\mathsf{C}^{-1})^\mathsf{T} \gamma^0
	= \minus (\psi^c)^\mathsf{T} (\mathsf{C}^{-1})^\mathsf{T} \gamma^0 \gamma^0
	= (\psi^c)^\mathsf{T} \mathsf{C}^{-1}
\end{align}
where we have used $\mathsf{C}^\mathsf{T} = \mathsf{C}^{-1} = \minus \mathsf{C}$ and $\mathsf{C}^{-1} \gamma^\mu \mathsf{C} = \minus (\gamma^\mu)^\mathsf{T}$. Hence, explicitly writing spinor indices $(a, b, \cdots)$ and using the Einstein summation convention, we can write
\begin{align}
	(\overline{N_\alpha})_a = (N_\alpha)_b (\mathsf{C}^{-1})_{ba}, \qquad
	(\overline{L_i^c})_a = (L_i)_b (\mathsf{C}^{-1})_{ba}, \qquad	
\end{align}
Suppressing flavor indices while keeping only spinor indices $(a, b, \cdots)$ and SU(2) indices $(i, j, \cdots)$, we can now rewrite the Lagrangian term as
\begin{align}
	\overline{N}_a (\widetilde{\phi}^\dag)^i \mathsf{L}_{ab} L_b^i &= N_c (\mathsf{C}^{-1})_{ca} (\widetilde{\phi}^*)^i \mathsf{L}_{ab} L_b^i
	= L_b^i (\widetilde{\phi}^*)^i \mathsf{L}_{ba} (\minus \mathsf{C}^{-1})_{ac} N_c \nonumber \\
	&= \minus L_b^i (\mathsf{C}^{-1})_{bd} \mathsf{C}_{de} (\widetilde{\phi}^*)^i \mathsf{L}_{ea} (\mathsf{C}^{-1})_{ac} N_c
	= \minus \overline{L^c}_d^i (\widetilde{\phi}^*)^i \mathsf{L}_{de} \mathsf{C}_{ea} (\mathsf{C}^{-1})_{ac} N_c \nonumber \\
	&= \minus \overline{L^c}_d^i (\widetilde{\phi}^*)^i \mathsf{L}_{dc} N_c.
\end{align}
We can also derive this identity in a more straightforward way, using $\mathcal{L}_1^\mathsf{T} = \mathcal{L}_1$ where $\mathcal{L}_1$ is an arbitrary Lagrangian term, as follows:
\begin{align}
	\overline{N_\alpha} \widetilde{\phi}^\dag \mathsf{L} L_i &= N_\alpha^\dag \gamma^0 \widetilde{\phi}^\dag \mathsf{L} L_i
	= (N_\alpha^\dag \gamma^0 \widetilde{\phi}^\dag \mathsf{L} L_i)^\mathsf{T}
	= L_i^\mathsf{T} \mathsf{L} \widetilde{\phi}^* (\gamma^0)^\mathsf{T} N_\alpha^* \nonumber \\
	&= (\gamma^0 \mathsf{C} L_i^*)^\dag \gamma^0 \mathsf{C} \mathsf{L} \widetilde{\phi}^* (\gamma^0)^\mathsf{T} \mathsf{C}^{-1} \gamma^0 (\gamma^0 \mathsf{C} N_\alpha^*)
	= \overline{L_i^c} \widetilde{\phi}^* \mathsf{L} \mathsf{C} (\gamma^0)^\mathsf{T} \mathsf{C}^{-1} \gamma^0 N_\alpha \nonumber \\
	&= \minus \overline{L_i^c} \widetilde{\phi}^* \mathsf{L} N_\alpha,
\end{align}
where we have used $\mathsf{C}^\dag = \mathsf{C}^{-1} = \minus \mathsf{C}$ and $\mathsf{C}^{-1} \gamma^\mu \mathsf{C} = \minus (\gamma^\mu)^\mathsf{T}$.

The correlation function for $L_i (x_1) \phi (x_1) \to L_j^c (x_2) \phi^* (x_2)$ is generally written as
\begin{align}
	\langle \Omega | T \big\{ \widetilde{\phi}^\mathsf{T} (x_2) L_j^c (x_2) \overline{L_i} (x_1) \widetilde{\phi} (x_1) \big\} | \Omega \rangle
	= \frac{\langle 0 | T \big\{ \widetilde{\phi}_I^\mathsf{T} (x_2) L_{I j}^c (x_2) \overline{L_{I i}} (x_1) \widetilde{\phi}_I (x_1) e^{i \int d^4 x \mathcal{L}_\text{int} [L_{I i}, \phi_I, N_{I \alpha}]} \big\} | 0 \rangle}{\langle 0 | T \big\{ e^{i \int d^4 x \mathcal{L}_\text{int} [L_{I i}, \phi_I, N_{I \alpha}]} \big\} | 0 \rangle},
\end{align}
where the subscript $I$ means the corresponding field is in the interaction picture. Now we drop $I$ for simplicity, and consider the numerator only with connected diagrams after disconnected ones are canceled out by the denominator. The tree-level contribution to this correlation function can be written as
\begin{align}
	&\int d^4 x \ d^4 y \ \langle 0 | T \big\{ \overline{L_j} (x_2) \mathsf{C}^{-1} \widetilde{\phi} (x_2) \overline{L_i} (x_1) \widetilde{\phi} (x_1) \big[ \minus i f_{j \alpha}^* \overline{N_\alpha} (x) \widetilde{\phi}^\dag (x) \mathsf{L} L_j (x) \big] \big[ \minus i f_{i \alpha}^* \overline{N_\alpha} (y) \widetilde{\phi}^\dag (y) \mathsf{L} L_i (y) \big] \big\} | 0 \rangle \nonumber \\
	&\ = \int d^4 x \ d^4 y \ \langle 0 | T \big\{ \widetilde{\phi}^\mathsf{T} (x_2) L_j^c (x_2) \overline{L_i} (x_1) \widetilde{\phi} (x_1) \big[ \minus i f_{j \alpha}^* \overline{L_j^c} (x) \widetilde{\phi}^* (x) \mathsf{L} N_\alpha (x) \big] \big[ \minus i f_{i \alpha}^* \overline{N_\alpha} (y) \widetilde{\phi}^\dag (y) \mathsf{L} L_i (y) \big] \big\} | 0 \rangle,
	\label{eq:Corr}
\end{align}
where equation \ref{eq:LTrick} have been used. Here, we have suppressed the SU(2) indices for simplicity, but the correlation function should be understood as an expression for each combination of SU(2) indices such as $(\widetilde{\phi}^\dag)^k (x) \mathsf{L} L_j^k (x)$ without the Einstein summation convention. In addition, to have legistimate contractions such as $L_j (x) \overline{L_j} (x_2)$ and $\widetilde{\phi} (x_2) \widetilde{\phi}^\dag (x)$ on the LH side of equation \ref{eq:Corr}, we have rewritten the external fields as
\begin{align}
	\widetilde{\phi}^\mathsf{T} L_j^c = (\widetilde{\phi}^\mathsf{T} L_j^c)^\mathsf{T}
	= (\gamma^0 \mathsf{C} L_j^*)^\mathsf{T} \widetilde{\phi}
	= \minus L_j^\dag \mathsf{C} (\gamma^0)^\mathsf{T} \widetilde{\phi}
	= \overline{L_j} \mathsf{C} \widetilde{\phi}
	= \minus \overline{L_j} \mathsf{C}^{-1} \widetilde{\phi},
\end{align}
where $\mathsf{C}^\mathsf{T} = \mathsf{C}^{-1} = \minus \mathsf{C}$ and $\mathsf{C}^{-1} \gamma^\mu \mathsf{C} = \minus (\gamma^\mu)^\mathsf{T}$ have been used. The minus sign due to this transformation has been absorbed into the first interaction term which had an opposite sign after applying the trick of equation \ref{eq:LTrick}. After this procedure, the contraction of $\overline{N_\alpha} (x) \overline{N_\alpha} (y)$ which would produce a Majorana-type propagator turned into the contraction of $N_\alpha (x) \overline{N_\alpha} (y)$ which would produce a Dirac-type propagator. The operator $\mathsf{C}$ in the Majorana-type propagator generated by the contraction of $\overline{N_\alpha} (x) \overline{N_\alpha} (y)$ would be canceled out by $\mathsf{C}^{-1}$ after $\overline{L_j}$. We can alternatively say that $\mathsf{C}$ has been absorbed into $L_j^c$ on the RH side of equation \ref{eq:Corr}. Treating $L_j^c$ as an ordinary field and using the contraction of $L_j^c (x) \overline{L_j^c} (x_2)$, we can simply write down the propagator of $L_j^c$ which is identical to that of $L_j$, without tediously tracking the spinor indices as in the presence of $\mathsf{C}$. The way of handling signs as we have discussed is the correct procedure since the LH and RH sides of equation \ref{eq:Corr} indeed have identical signs after contraction of all the fields. In other words, the sign change due to the transformation of interaction terms as in equation \ref{eq:LTrick} does not really cause any sign change such as $\minus i f_{i \alpha} \to i f_{i \alpha}$ in Feynman rules.

For the tree-level scattering $L_i^k \widetilde{\phi}^k \to (L_j^c)^l (\widetilde{\phi}^*)^l$ where $k$ and $l$ are the SU(2) indices, the associated interaction terms are $f_{i \alpha}^* \overline{L_i^c} \widetilde{\phi}^* \mathsf{L} N_\alpha$ and $\minus f_{j \alpha}^* \overline{N_\alpha} \widetilde{\phi}^\dag \mathsf{L} L_j$. In terms of a Dirac-type propagator, the scattering amplitude is written as
\begin{align}
	i \mathcal{M}_\text{tree} \big[ L_i^k \widetilde{\phi}^k \to (L_j^c)^l (\widetilde{\phi}^*)^l \big] &= \sum_\alpha \overline{u_{L_j^l}^c} (\textbf{p}_{L_j^l}) (\minus i f_{j \alpha}^* \mathsf{L}) \frac{i}{\slashed{p} - m_{N_\alpha}} (\minus i f_{i \alpha}^* \mathsf{L}) u_{L_i^k} (\textbf{p}_{L_i^k}) \nonumber \\
	&= \sum_\alpha \overline{v_{L_j^l}} (\textbf{p}_{L_j^l}) (\minus i f_{j \alpha}^* \mathsf{L}) \frac{i}{\slashed{p} - m_{N_\alpha}} (\minus i f_{i \alpha}^* \mathsf{L}) u_{L_i^k} (\textbf{p}_{L_i^k}).
\end{align}
Here, we have used $u_s^c = v_{-s}$, where $s$ denotes spin and it has been suppressed above for simplicity.
\begin{figure}[h]
	\centering
	\subfloat[]{
		\includegraphics[width = 60 mm]{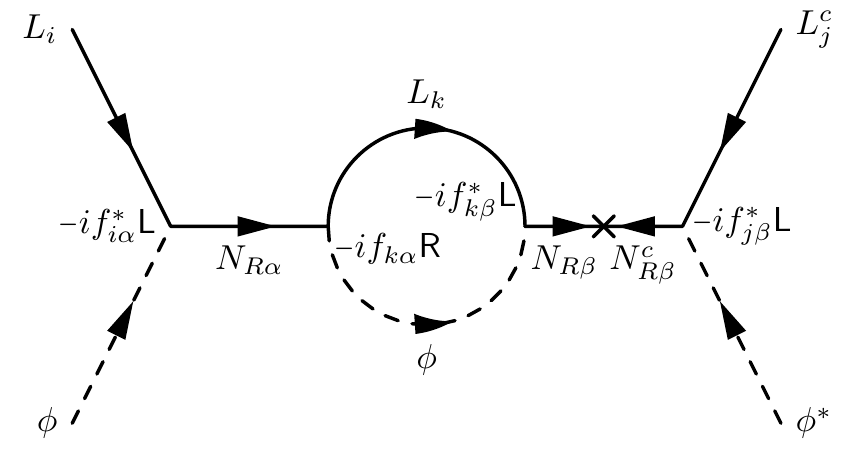}
		\label{fig:LPLcPcLoopR}
	} \qquad \quad
	\subfloat[]{
		\includegraphics[width = 60 mm]{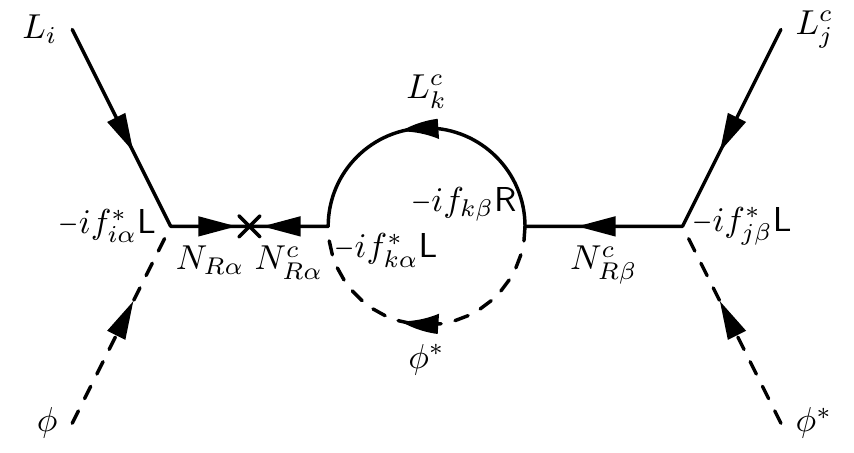}
		\label{fig:LPLcPcLoopL}
	}
	\caption{One-loop corrections to the propagator in $L_i \phi \to L_j^c \phi^*$.}
	\label{fig:LPLcPcLoop}
\end{figure}
As the second example, let us consider one-loop corrections to the propagator in $L_i \phi \to L_j^c \phi^*$ whose diagrams are given in figure \ref{fig:LPLcPcLoop}. For figure \ref{fig:LPLcPcLoopR}, we can apply the trick to the fields in the correlation function as follows:
\begin{align}
	&\overline{L_j} \mathsf{C}^{-1} \widetilde{\phi} \overline{L_i} \widetilde{\phi}
		(\minus i f_{j \beta}^* \overline{N_\beta} \widetilde{\phi}^\dag \mathsf{L} L_j) (\minus i f_{k \beta}^* \overline{N_\beta} \widetilde{\phi}^\dag \mathsf{L} L_k) (\minus i f_{k \alpha} \overline{L_k} \widetilde{\phi} \mathsf{R} N_\alpha) (\minus i f_{i \alpha}^* \overline{N_\alpha} \widetilde{\phi}^\dag \mathsf{L} L_i) \nonumber \\
	&\to \minus \widetilde{\phi}^\mathsf{T} L_j^c \overline{L_i} \widetilde{\phi}
		(i f_{j \beta}^* \overline{L_j^c} \widetilde{\phi}^* \mathsf{L} N_\beta) (\minus i f_{k \beta}^* \overline{N_\beta} \widetilde{\phi}^\dag \mathsf{L} L_k) (\minus i f_{k \alpha} \overline{L_k} \widetilde{\phi} \mathsf{R} N_\alpha) (\minus i f_{i \alpha}^* \overline{N_\alpha} \widetilde{\phi}^\dag \mathsf{L} L_i) \nonumber \\
	&\to \widetilde{\phi}^\mathsf{T} L_j^c \overline{L_i} \widetilde{\phi}
		(\minus i f_{j \beta}^* \overline{L_j^c} \widetilde{\phi}^* \mathsf{L} N_\beta) (\minus i f_{k \beta}^* \overline{N_\beta} \widetilde{\phi}^\dag \mathsf{L} L_k) (\minus i f_{k \alpha} \overline{L_k} \widetilde{\phi} \mathsf{R} N_\alpha) (\minus i f_{i \alpha}^* \overline{N_\alpha} \widetilde{\phi}^\dag \mathsf{L} L_i),
\end{align}
where the sign change in the first interaction term is again compensated. The scattering amplitude is given by
\begin{align}
	&i \mathcal{M}_\text{1-loop}^R \big[ L_i^k \widetilde{\phi}^k \to (L_j^c)^l (\widetilde{\phi}^*)^l \big] \nonumber \\
	&\qquad = \sum_{\alpha, \beta} \overline{v_{L_j^l}} (\textbf{p}_{L_j^l}) (\minus i f_{j \beta}^* \mathsf{L}) \frac{i}{\slashed{p} - m_{N_\beta}} \big[ i \slashed{p} \mathsf{R} (\Sigma_R)_{\beta \alpha} (p^2) \big] \frac{i}{\slashed{p} - m_{N_\alpha}} (\minus i f_{i \alpha}^* \mathsf{L}) u_{L_i^k} (\textbf{p}_{L_i^k}).
\end{align}
In addition, for figure \ref{fig:LPLcPcLoopL}, the sign changes in the first three interaction terms are also compensated as follows:
\begin{align}
	&\overline{L_j} \mathsf{C}^{-1} \widetilde{\phi} \overline{L_i} \widetilde{\phi}
		(\minus i f_{j \beta}^* \overline{N_\beta} \widetilde{\phi}^\dag \mathsf{L} L_j) (\minus i f_{k \beta} \overline{L_k} \widetilde{\phi} \mathsf{R} N_\beta) (\minus i f_{k \alpha}^* \overline{N_\alpha} \widetilde{\phi}^\dag \mathsf{L} L_k) (\minus i f_{i \alpha}^* \overline{N_\alpha} \widetilde{\phi}^\dag \mathsf{L} L_i) \nonumber \\
	&\to \minus \widetilde{\phi}^\mathsf{T} L_j^c \overline{L_i} \widetilde{\phi}
		(i f_{j \beta}^* \overline{L_j^c} \widetilde{\phi}^* \mathsf{L} N_\beta) (i f_{k \beta} \overline{N_\beta} \widetilde{\phi}^\mathsf{T} \mathsf{R} L_k^c) (i f_{k \alpha}^* \overline{L_k^c} \widetilde{\phi}^* \mathsf{L} N_\alpha) (\minus i f_{i \alpha}^* \overline{N_\alpha} \widetilde{\phi}^\dag \mathsf{L} L_i) \nonumber \\
	&\to \widetilde{\phi}^\mathsf{T} L_j^c \overline{L_i} \widetilde{\phi}
		(\minus i f_{j \beta}^* \overline{L_j^c} \widetilde{\phi}^* \mathsf{L} N_\beta) (\minus i f_{k \beta} \overline{N_\beta} \widetilde{\phi}^\mathsf{T} \mathsf{R} L_k^c) (\minus i f_{k \alpha}^* \overline{L_k^c} \widetilde{\phi}^* \mathsf{L} N_\alpha) (\minus i f_{i \alpha}^* \overline{N_\alpha} \widetilde{\phi}^\dag \mathsf{L} L_i).
\end{align}
and thus
\begin{align}
	&i \mathcal{M}_\text{1-loop}^L \big[ L_i^k \widetilde{\phi}^k \to (L_j^c)^l (\widetilde{\phi}^*)^l \big] \nonumber \\
	&\qquad = \sum_{\alpha, \beta} \overline{v_{L_j^l}} (\textbf{p}_{L_j^l}) (\minus i f_{j \beta}^* \mathsf{L}) \frac{i}{\slashed{p} - m_{N_\beta}} \big[ i \slashed{p} \mathsf{L} (\Sigma_R^\mathsf{T})_{\beta \alpha} (p^2) \big] \frac{i}{\slashed{p} - m_{N_\alpha}} (\minus i f_{i \alpha}^* \mathsf{L}) u_{L_i^k} (\textbf{p}_{L_i^k}).
\end{align}

In short, we can always replace a Majorana-type propagator with a Dirac-type propagator. The sign change of Yukawa couplings due to the application of the trick does not require any change of Feynman rules.

\section{One-loop contributions to the self-energy and vertex}		\label{sec:1Loop}
In this section, we explicitly calculate the one-loop diagrams used in this paper, applying the trick discussed in appendix \ref{sec:TrickMaj} when it is required. For simplicity, we assume that $L_i$ and $\phi$ are massless.

\subsection*{B.1~~~Self-energy}
\begin{figure}[h]
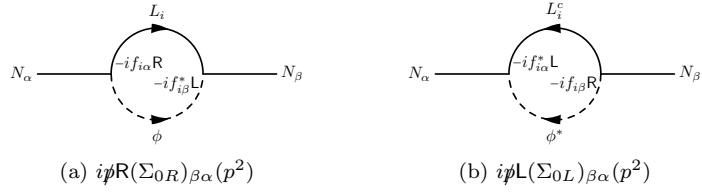

	\centering
	\subfloat[$i \slashed{p} \mathsf{R} (\Sigma_{0 R})_{\beta \alpha} (p^2)$]{
		\includegraphics[width = 40 mm]{SER}
		\label{fig:SERLoop}
	} \qquad \quad
	\subfloat[$i \slashed{p} \mathsf{L} (\Sigma_{0 L})_{\beta \alpha} (p^2)$]{
		\includegraphics[width = 40 mm]{SEL}
		\label{fig:SELLoop}
	}
	\caption{One-loop diagrams which contribute to the self-energy of $N_\alpha$.}
	\label{fig:SELoop}
\end{figure}

This is an elementary one-loop calculation, but it is presented for completeness. Initially we will assume $p^2$ is real, but later it will be analytically continued to a complex value. The one-loop diagrams are given in figure \ref{fig:SELoop}.
Considering two sets of degrees of freedom in $L_i (k) \phi (p - k)$ due to two pairs of SU(2) components, we can write
\begin{align}
	i \slashed{p} \mathsf{R} (\Sigma_{0 R})_{\beta \alpha} (p^2)
	&= 2 \sum_i \int \frac{d^d k}{(2 \pi)^d} \ (\minus i f_{i \beta}^* \mathsf{L}) \frac{i}{\slashed{k}} (\minus i f_{i \alpha} \mathsf{R}) \frac{i}{(p - k)^2}
	= 2 \mathsf{L} \sum_i f_{i \beta}^* f_{i \alpha} \int \frac{d^d k}{(2 \pi)^d} \ \frac{\slashed{k}}{k^2 (p - k)^2} \nonumber \\
	&= 2 \mathsf{L} \sum_i f_{i \beta}^* f_{i \alpha} \int \frac{d^d k}{(2 \pi)^d} \int_0^1 dx \ \frac{\slashed{k}}{[(1 - x) k^2 + x (p - k)^2 ]^2},
\end{align}
where $d = 4 - 2 \epsilon$ is the spacetime dimension. Manipulating the Feynman parameter and the momentum integration in the standard way, we rewrite
\begin{align}
	\int \frac{d^d k}{(2 \pi)^d} \ \frac{\slashed{k}}{[(1 - x) k^2 + x (p - k)^2 ]^2} = \int \frac{d^d \ell}{(2 \pi)^d} \ \frac{x \slashed{p}}{[\ell^2 + x (1 - x) p^2]^2},
\end{align}
where $\ell = k - x p$. Hence,
\begin{align}
	\int &\frac{d^d k}{(2 \pi)^d} \ \frac{\slashed{k}}{k ^2 (p - k)^2}
	= \int_0^1 dx \int \frac{d^d \ell}{(2 \pi)^d} \ \frac{x \slashed{p}}{[\ell^2 + x (1 - x) p^2]^2} \nonumber \\
	&= i \slashed{p} \int_0^1 dx \ x \frac{\Gamma (2 - \frac{d}{2})}{(4 \pi)^{d/2} \Gamma (2)} [-x (1 - x) p^2]^{-(2 - d/2)}
	= i \slashed{p} \int_0^1 dx \ x \frac{1}{(4 \pi)^2} (4 \pi)^\epsilon \Gamma (\epsilon) [-x (1 - x) p^2]^{\minus \epsilon} \nonumber \\
	&= i \slashed{p} \frac{1}{(4 \pi)^2} \int_0^1 dx \ x \left[ \frac{1}{\epsilon} - \gamma + \log{4 \pi} - \log \{-x (1 - x) p^2\} + \mathcal{O} (\epsilon) \right] \nonumber \\
	&= i \slashed{p} \frac{1}{32 \pi^2} \left[ \frac{1}{\epsilon} - \gamma + \log{4 \pi} + 2 - \log{p^2} + i \pi + \mathcal{O} (\epsilon) \right],
\end{align}
where we have used $\log{(\minus 1)} = \minus i \pi$. So far $p^2$ has been assumed to be real. The value of $p^2$ as a complex number is supposed to be in the region around $(0, \infty)$ on the real axis in the complex plane, because a physical pole is generally given by $p_{\widehat{N}_{\widehat{\alpha}}}^2 = m_{\widehat{N}_{\widehat{\alpha}}}^2 - i m_{\widehat{N}_{\widehat{\alpha}}} \Gamma_{\widehat{N}_{\widehat{\alpha}}}$ with $m_{\widehat{N}_{\widehat{\alpha}}} \gg \Gamma_{\widehat{N}_{\widehat{\alpha}}}$. Choosing the branch cut of $\log{z}$ at $(\minus \infty, 0]$ so that it is an analytic function except for a region around $(\minus \infty, 0]$, we may analytically continue $\log{p^2}$ to allow a complex value for $p^2$ such as $p_{\widehat{N}_{\widehat{\alpha}}}^2 = m_{\widehat{N}_{\widehat{\alpha}}}^2 - i m_{\widehat{N}_{\widehat{\alpha}}} \Gamma_{\widehat{N}_{\widehat{\alpha}}}$. Now we can write
\begin{align}
	\int \frac{d^d k}{(2 \pi)^d} \ \frac{\slashed{k}}{k^2 (p - k)^2} &= i \slashed{p} \frac{1}{32 \pi^2} \left[ \frac{1}{\epsilon} - \gamma + \log{4 \pi} + 2 - \log{|p^2|} + i (\pi - \arg{[p^2]}) + \mathcal{O} (\epsilon) \right].
\end{align}
Defining
\begin{align}
	\log{\Lambda^2} \coloneqq \frac{1}{\epsilon} - \gamma + \log{4 \pi} + 2,
\end{align}
we finally obtain
\begin{align}
	(\Sigma_{0 R})_{\beta \alpha} (p^2) = \sum_i \frac{f_{i \beta}^* f_{i \alpha}}{16 \pi^2} \bigg[ \minus \log{\bigg( \frac{|p^2|}{\Lambda^2} \bigg)} + i (\pi - \arg{[p^2]}) \bigg].
\end{align}
In addition, it is also straightforward to show that
\begin{align}
	\Sigma_{0 L} (p^2) = \Sigma_{0 R}^\mathsf{T} (p^2),
\end{align}
which completes the calculation of the bare self-energy for complex-valued $p^2$.

\subsection*{B.2~~~Vertex}
\begin{figure}[h]
	\centering
	\subfloat[$i \mathsf{R} \slashed{p} (V_{RL})_{i \alpha} (p^2)$]{
		\includegraphics[width = 34 mm]{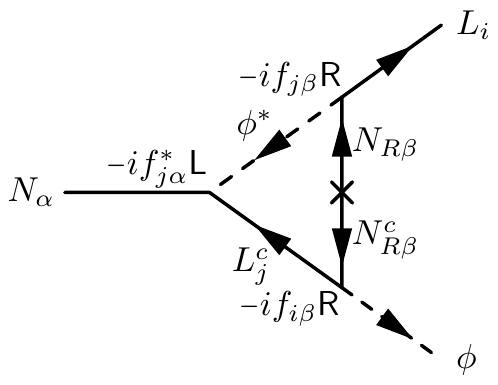}
		\label{fig:NLPLoopVM}
	} \qquad \quad
	\subfloat[$-i \mathsf{L} \slashed{p} (V_{LR})_{i \alpha} (p^2)$]{
		\includegraphics[width = 34 mm]{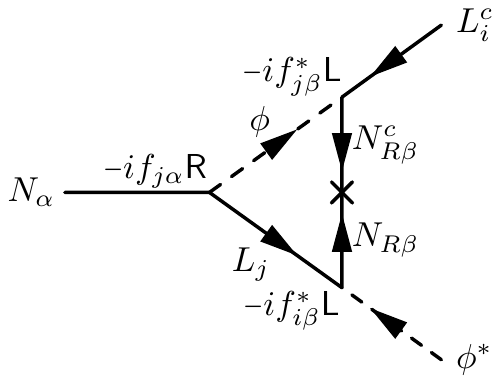}
		\label{fig:NLcPcLoopVM}
	}
	\caption{One-loop diagrams which contribute to the vertex.}
	\label{fig:V1Loop}
\end{figure}

Here, we calculate the one-loop correction to the vertex, whose Feynman diagrams are given in figure \ref{fig:V1Loop}.
Note that there does not exist any diagram with a Dirac-type propagator of $N_\beta$ in the vertex-loop. In those diagrams, the Majorana propagators are not in the $s$-channel, and thus it is unclear whether the resummed propagator should be used or not since the associated non-perturbative effect emerges around the resonance. Even though we decide to use the resummed propagator, there still exists an ambiguity since the final form of the propagator after the pole expansion is only valid around the resonance. For an off-shell Majorana particle, the neglected part in the propagator after the pole expansion could be comparable to the dominant part we have kept. Hence, for a careful calculation with the resummed propagator, we have to check whether the contribution of the neglected part matters or not up to the working precision. Here, we do not consider such a subtlety, and use only the tree-level ones for those internal propagators for simplicity.

We introduce the vertex function $V (\slashed{p})$ and decompose it into chiral components
\begin{align}
	V (\slashed{p}) = \mathsf{R} V_{RR} (p^2) + \mathsf{R} \slashed{p} V_{RL} (p^2) + \mathsf{L} \slashed{p} V_{LR} (p^2) + \mathsf{L} V_{LL} (p^2),
\end{align}
where $V_{RR} (p^2) = f$ and $V_{LL} (p^2) = f^*$ due to the absence of diagrams with Dirac-type propagators. For example, $V (\slashed{p})$ contributes to $L_i^k \phi^k \to L_j^l \phi^l$ where $k$ and $l$ are SU(2) indices as follows:
\begin{align}
	i &\mathcal{M} (L_i^k \phi^k \to L_j^l \phi^l) \nonumber \\
	&= \sum_{\alpha, \beta} \overline{u_{L_j^l}} (\mathbf{p}_{L_j^l}) \mathsf{R} \big\{ \big[ \minus i V_{j \beta} (\slashed{p}) \big] \big[ i \Delta_{\beta \alpha} (\slashed{p}) \big] \big[ \minus i V_{i \alpha} (\slashed{p}) \big] \big\} \mathsf{L} u_{L_i^k} (\mathbf{p}_{L_i^k}) \nonumber \\
	&= \sum_{\alpha, \beta} \overline{u_{L_j^l}} (\mathbf{p}_{L_j^l}) \mathsf{R} \bigg\{ (\minus i f_{j \beta}) \big[ i \Delta_{RR} (p^2) \big]_{\beta \alpha} \big[ \minus i \slashed{p} V_{RL} (p^2) \big]_{i \alpha}
		+ (\minus i f_{j \beta}) \big[ i \slashed{p} \Delta_{RL} (p^2) \big]_{\beta \alpha} (\minus i f_{i \alpha}^*) \nonumber \\
		&\qquad \qquad + \big[ \minus i \slashed{p} V_{RL} (p^2) \big]_{j \beta} \big[ i \slashed{p} \Delta_{LR} (p^2) \big]_{\beta \alpha} \big[ \minus i \slashed{p} V_{RL} (p^2) \big]_{i \alpha}
		+ \big[ \minus i \slashed{p} V_{RL} (p^2) \big]_{j \beta} \big[ i \Delta_{LL} (p^2) \big]_{\beta \alpha} (\minus i f_{i \alpha}^*) \bigg\} \nonumber \\
		&\qquad \quad u_{L_i^k} (\mathbf{p}_{L_i^k}).
\end{align}
First we calculate $V_{RL} (p^2)$ for $N_\alpha (p) \to L_j^l (k) \phi^l (p - k)$:
\begin{align}
	\minus i \mathsf{R} \slashed{p} (V_{RL})_{i \alpha} (p^2) &= \sum_{j, \beta} \int \frac{d^d \ell}{(2 \pi)^d} (\minus i f_{i \beta} \mathsf{R}) \frac{i}{\slashed{\ell} - m_{N_\beta}} (\minus i f_{j \beta} \mathsf{R}) \frac{i}{\slashed{\ell} + \slashed{p} - \slashed{k}} (\minus i f_{j \alpha}^* \mathsf{L}) \frac{i}{(k - \ell)^2} \nonumber \\
	&= \mathsf{R} \sum_{j, \beta} f_{i \beta} f_{j \beta} f_{j \alpha}^* \int \frac{d^d \ell}{(2 \pi)^d} \frac{m_{N_\beta} (\slashed{\ell} + \slashed{p} - \slashed{k})}{(\ell^2 - m_{N_\beta}^2) (\ell + p - k)^2 (k - \ell)^2} \nonumber \\
	&= 2 \mathsf{R} \sum_{j, \beta} f_{i \beta} f_{j \beta} f_{j \alpha}^* \int \frac{d^d \ell}{(2 \pi)^d} \int_0^1 dx \int_0^1 dy \int_0^1 dz \ \delta (1 - x - y - z) \nonumber \\
		&\qquad \qquad \qquad \qquad \qquad \frac{m_{N_\beta} (\slashed{\ell} + \slashed{p} - \slashed{k})}{[x (\ell^2 - m_{N_\beta}^2) + y (\ell + p - k)^2 + z (k - \ell)^2]^3}.
\end{align}
The integration over $\ell$ can be rewritten as
\begin{align}
	\int \frac{d^d \ell}{(2 \pi)^d} &\frac{m_{N_\beta} (\slashed{\ell} + \slashed{p} - \slashed{k})}{[x (\ell^2 - m_{N_\beta}^2) + y (\ell + p - k)^2 + z (k - \ell)^2]^3}
	= \int \frac{d^d q}{(2 \pi)^d} \frac{m_{N_\beta} [\slashed{q} + (1 - y) (\slashed{p} - \slashed{k}) + z \slashed{k}]}{[q^2 + 2 y z (p \cdot k) - x m_{N_\beta}^2]^3} \nonumber \\
	&\to \int \frac{d^d q}{(2 \pi)^d} \frac{(1 - y) m_{N_\beta} \slashed{p}}{(q^2 + y z p^2 - x m_{N_\beta}^2)^3},
\end{align}
where $q = \ell + y (p - k) - z k$ and we have used $2 p \cdot k = 2 (p - k) \cdot k = p^2$ and $\overline{u_{L_i}} \slashed{k} = 0$. Defining $r_\beta (p^2) \coloneqq m_{N_\beta}^2 / p^2$,
we write
\begin{align}
	\int_0^1 &dx \int_0^1 dy \int_0^1 dz \ \delta (1 - x - y - z) \int \frac{d^d q}{(2 \pi)^d} \ \frac{(1 - y) m_{N_\beta} \slashed{p}}{(q^2 + y z p^2 - x m_{N_\beta}^2)^3} \nonumber \\
	&= i \int_0^1 dx \int_0^1 dy \int_0^1 dz \ \delta (1 - x - y - z) \ (1 - y) m_{N_\beta} \slashed{p} \bigg[ \frac{1}{2 (4 \pi)^2} (-y z p^2 + x m_{N_\beta}^2)^{-1} + \mathcal{O} (\epsilon) \bigg] \nonumber \\
	&= \frac{i \sqrt{r_\beta}}{2 (4 \pi)^2} \frac{\slashed{p}}{|p|} \int_0^1 dz \int_0^{1 - z} dy \ \bigg[ \frac{1 - y}{-y z + r_\beta (1 - y - z)} + \mathcal{O} (\epsilon) \bigg] \nonumber \\
	&= \frac{i \sqrt{r_\beta}}{2 (4 \pi)^2} \frac{\slashed{p}}{|p|} \int_0^1 dz \ \bigg[ \frac{1 - z}{r_\beta + z} - \frac{(1 + r_\beta) z}{(r_\beta + z)^2} \log{\bigg( \frac{z}{r_\beta} \bigg)} + i \pi \frac{(1 + r_\beta) z}{(r_\beta + z)^2} + \mathcal{O} (\epsilon) \bigg],
\end{align}
where we have used $\log{(\minus 1)} = \minus i \pi$ in the last step. Using
\begin{align}
	&\int_0^1 dz \ \frac{z}{(r_\beta + z)^2} = \log{(1 + r_\beta^{-1})} - \frac{1}{1 + r_\beta}, \\
	&s_\beta (p^2) \coloneqq \int_0^1 dz \ \left[ \frac{1 - z}{r_\beta + z} - \frac{(1 + r_\beta) z}{(r_\beta + z)^2} \log{\left( \frac{z}{r_\beta} \right)} \right],
\end{align}
we obtain
\begin{align}
	\int &\frac{d^d \ell}{(2 \pi)^d} \int_0^1 dx \int_0^1 dy \int_0^1 dz \ \delta (1 - x - y - z)
		\frac{m_{N_\beta} (\slashed{\ell} + \slashed{p} - \slashed{k})}{[x (\ell^2 - m_{N_\beta}^2) + y (\ell + p - k)^2 + z (k - \ell)^2]^3} \nonumber \\
	&\to \int_0^1 dx \int_0^1 dy \int_0^1 dz \ \delta (1 - x - y - z) \ (\minus i) \int \frac{d^d q_E}{(2 \pi)^d} \frac{(1 - y) m_{N_\beta} \slashed{p}}{(q_E^2 - y z p^2 + x m_{N_\beta}^2)^3} \nonumber \\
		&= \minus \frac{i \sqrt{r_\beta}}{2 (4 \pi)^2} \frac{\slashed{p}}{|p|} \big[ s_\beta - i \pi \left\{1 - (1 + r_\beta) \log{(1 + r_\beta^{-1})} \right\} + \mathcal{O} (\epsilon) \big].
\end{align}
Hence,
\begin{empheq}[box=\fbox]{align}
	\minus i &\slashed{p} (V_{RL})_{i \alpha} (p^2) \nonumber \\
	&= \minus i \slashed{p} \sum_\beta \frac{f_{i \beta} (f^\mathsf{T} f^*)_{\beta \alpha}}{16 \pi^2} \frac{m_{N_\beta}}{p^2} \Bigg[ s_\beta (p^2) - i \pi \bigg\{ 1 - \bigg( 1 + \frac{m_{N_\beta}^2}{p^2} \bigg) \log{\bigg( 1 + \frac{p^2}{m_{N_\beta}^2} \bigg)} \bigg\} \Bigg],
\end{empheq}
and similarly
\begin{empheq}[box=\fbox]{align}
	\minus i &\slashed{p} (V_{LR})_{i \alpha} (p^2) \nonumber \\
	&= \minus i \slashed{p} \sum_\beta \frac{f_{i \beta}^* (f^\dag f)_{\beta \alpha}}{16 \pi^2} \frac{m_{N_\beta}}{p^2} \Bigg[ s_\beta (p^2) - i \pi \bigg\{ 1 - \bigg( 1 + \frac{m_{N_\beta}^2}{p^2} \bigg) \log{\bigg( 1 + \frac{p^2}{m_{N_\beta}^2} \bigg)} \bigg\} \Bigg],
\end{empheq}
where to repeat for clearance
\begin{align}
	s_\beta (p^2) \coloneqq \int_0^1 dx \ \bigg[ \frac{1 - x}{r_\beta + x} - \frac{(1 + r_\beta) x}{(r_\beta + x)^2} \log{\left( \frac{x}{r_\beta} \right)} \bigg], \qquad
	r_\beta (p^2) \coloneqq \frac{m_{N_\beta}^2}{p^2}.
\end{align}
We can define momentum-dependent mixing matrices:
\begin{empheq}[box=\fbox]{align}
	(D_R)_{ji} (\slashed{p}) &\coloneqq \delta_{ji} + \slashed{p} \sum_\alpha \frac{f_{j \alpha}^* f_{i \alpha}^*}{16 \pi^2} \frac{m_{N_\alpha}}{p^2} \Bigg[ s_\alpha (p^2) - i \pi \bigg\{ 1 - \bigg( 1 + \frac{m_{N_\alpha}^2}{p^2} \bigg) \log{\bigg( 1 + \frac{p^2}{m_{N_\alpha}^2} \bigg)} \bigg\} \Bigg], \\
	(D_L)_{ji} (\slashed{p}) &\coloneqq \delta_{ji} + \slashed{p} \sum_\alpha \frac{f_{j \alpha} f_{i \alpha}}{16 \pi^2} \frac{m_{N_\alpha}}{p^2} \Bigg[ s_\alpha (p^2) - i \pi \bigg\{ 1 - \bigg( 1 + \frac{m_{N_\alpha}^2}{p^2} \bigg) \log{\bigg( 1 + \frac{p^2}{m_{N_\alpha}^2} \bigg)} \bigg\} \Bigg],
\end{empheq}
so that
\begin{align}
	f + \slashed{p} V_{RL} (p^2) = D_R (\slashed{p}) f, \qquad
	f^* + \slashed{p} V_{LR} (p^2) = D_L (\slashed{p}) f^*.
\end{align}
Furthermore, choosing the branch cut of $\log{z}$ at $(\minus \infty, 0]$ on the real axis, we may analytically continue $\log{p^2}$ and $\log{(1 + p^2 / m_{N_\alpha}^2)}$ to allow a complex value for $p^2$. \\

The effective Yukawa couplings incorporating all the loop effects in the vertex as well as in the field-strength can be defined by
\begin{align}
	\boxed{\widehat{F}_{i \widehat{\alpha}} \coloneqq (D_R^{\widehat{N}_{\widehat{\alpha}}} f C_R^{\widehat{N}_{\widehat{\alpha}}})_{i \alpha}, \qquad
	\widehat{F}^c_{i \widehat{\alpha}} \coloneqq (D_L^{\widehat{N}_{\widehat{\alpha}}} f^* C_L^{\widehat{N}_{\widehat{\alpha}}})_{i \alpha},}
	\label{eq:Feff}
\end{align}
where
\begin{align}
	\boxed{D_R^{\widehat{N}_{\widehat{\alpha}}} \coloneqq D_R (p_{\widehat{N}_{\widehat{\alpha}}}), \qquad
	D_L^{\widehat{N}_{\widehat{\alpha}}} \coloneqq D_L (p_{\widehat{N}_{\widehat{\alpha}}}).}
\end{align}


\end{document}